%% file: main.tex
\documentclass[conference]{IEEEtran}
\IEEEoverridecommandlockouts
\usepackage[sort,nocompress]{cite}
\usepackage{amsmath,amssymb,amsfonts}
\usepackage{algorithm}
\usepackage{algpseudocode}
\usepackage{multirow}
\usepackage{makecell}
\usepackage{array}
\newcolumntype{M}[1]{>{\raggedright\arraybackslash}m{#1}}
\usepackage{enumerate}
\usepackage{graphicx}
\usepackage{textcomp}
\usepackage{xcolor}
\def\BibTeX{{\rm B\kern-.05em{\sc i\kern-.025em b}\kern-.08em
    T\kern-.1667em\lower.7ex\hbox{E}\kern-.125emX}}
\usepackage{braket}
\usepackage[affil-it]{authblk}

\usepackage[bookmarks=true,breaklinks=true,colorlinks,linkcolor=blue,citecolor=blue,urlcolor=black]{hyperref}

\begin{document}

\title{\fontsize{23}{28}\selectfont{Systematic Crosstalk Mitigation for Superconducting Qubits via Frequency-Aware Compilation}}

\author[]{Yongshan Ding\thanks{Corresponding author: yongshan@uchicago.edu}, Pranav Gokhale, Sophia Fuhui Lin \\ Richard Rines, Thomas Propson, and Frederic T. Chong}

\affil{Department of Computer Science, University of Chicago, Chicago, IL 60615, USA}


\maketitle

\begin{abstract}
One of the key challenges in current Noisy Intermediate-Scale Quantum (NISQ) computers is to control a quantum system with high-fidelity quantum gates. There are many reasons a quantum gate can go wrong -- for superconducting transmon qubits in particular, one major source of gate error is the unwanted crosstalk between neighboring qubits due to a phenomenon called frequency crowding. We motivate a systematic approach for understanding and mitigating the crosstalk noise when executing near-term quantum programs on superconducting NISQ computers. We present a general software solution to alleviate frequency crowding by systematically tuning qubit frequencies according to input programs, trading parallelism for higher gate fidelity when necessary. The net result is that our work dramatically improves the crosstalk resilience of tunable-qubit, fixed-coupler hardware, matching or surpassing other more complex architectural designs such as tunable-coupler systems. On NISQ benchmarks, we improve worst-case program success rate by 13.3x on average, compared to existing traditional serialization strategies.  
\end{abstract}


\input{introduction}

\input{background}
\input{related}
\input{frequency}

\input{approach} 
\input{evaluation}
\input{results}

\input{validation}

\input{conclusion}

\section*{Acknowledgments}
This work is funded in part by EPiQC, an NSF Expedition in Computing, under grants CCF-1730449; in part by STAQ, under grant NSF Phy-1818914; and in part by DOE grants DE-SC0020289 and DE-SC0020331. This work was completed in part with resources provided by the University of Chicago Research Computing Center. PG is supported by the Department of Defense (DoD) through the National Defense Science $\&$ Engineering Graduate Fellowship (NDSEG) Program. We thank Kenneth Brown, Ike Chuang, Morten Kjaergaard, Nelson Leung, Prakash Murali, David Schuster, and Christopher Wang for fruitful discussions. We also thank the anonymous reviewers for their valuable comments and suggestions.

\appendices
\input{Append}

\bibliographystyle{IEEEtranS}
\bibliography{ref}

\end{document}

%% file: introduction.tex
\section{Introduction}\label{sec:intro}

Current Noisy Intermediate-Scale Quantum (NISQ) computers \cite{preskill2018quantum, arute2019quantum, wright2019benchmarking, IBM, Intel}  aim to isolate and control a non-trivial quantity of quantum bits (qubits) with high precision. Scaling  up  a  quantum  computer requires  improvements in both the quality of qubits (with longer lifetime) and the quality of gates (with higher fidelity).  


In case of superconducting 
transmon qubits \cite{koch2007charge, hutchings2017tunable, barends2013coherent}, which is the subject of this work, gate speeds have been achieved three to four orders of magnitude faster than qubit lifetime \cite{barends2019diabatic, corcoles2015demonstration, reed2012realization, barends2016digitized}. Although fast gates are desirable; they are prone to errors caused by imprecise control. Among all sources of gate errors, crosstalk is the most dominant \cite{mundada2019suppression, mckay2019three}. Errors caused by crosstalk, such as exchange of excitation and leakage to non-computational states, are found to have detrimental effect to quantum states, and such errors can accumulate as we execute a program \cite{barends2019diabatic}.

\begin{figure}[t]
    \centering
    \includegraphics[width=0.8\linewidth, trim= -1cm 0cm 1cm 0]{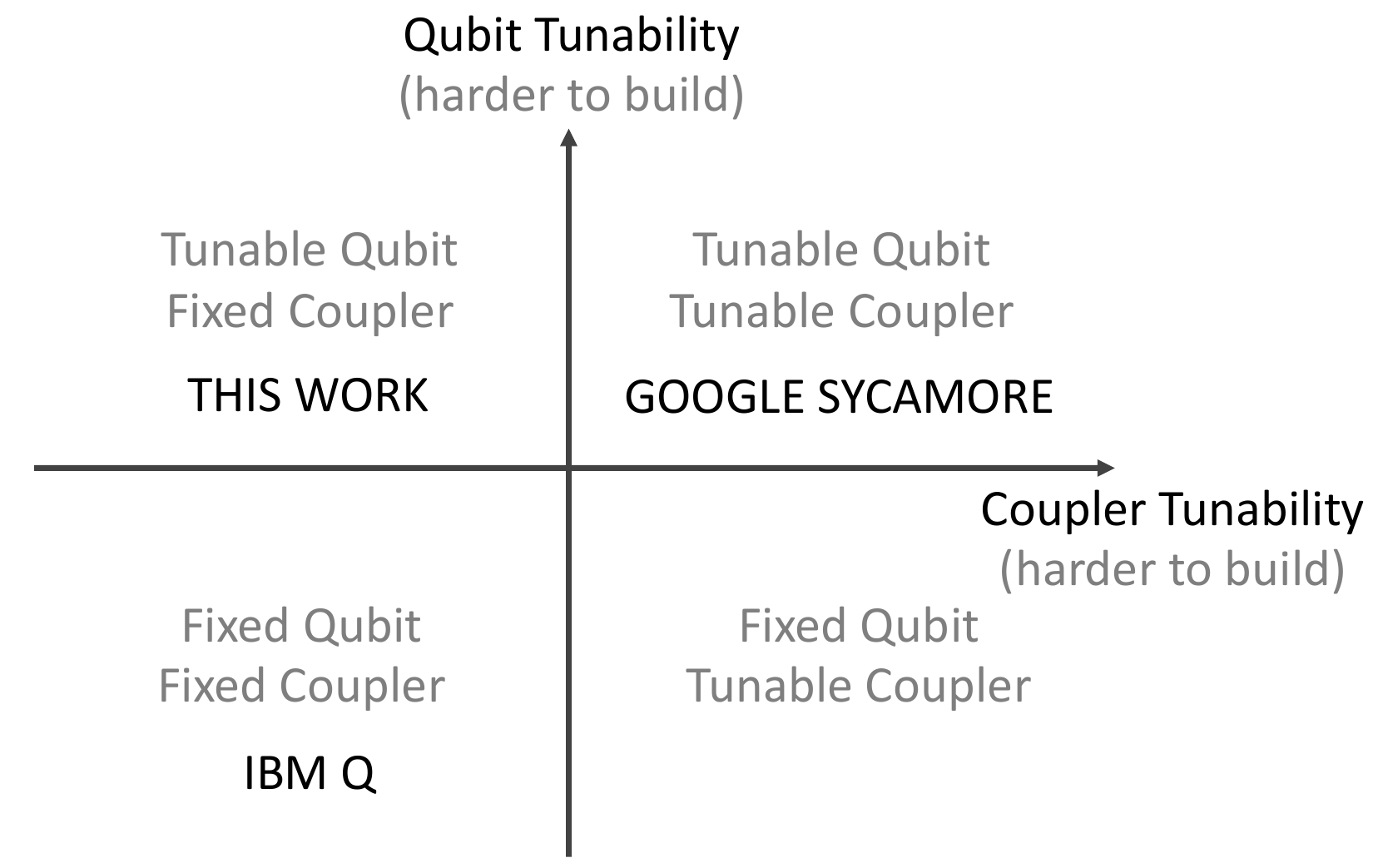}
    \caption{Technological design choices for mitigating crosstalk. Higher tunability offers better control over the device, but induces higher fabrication overhead and sensitivity to control noise. Our work targets a balanced design, i.e. tunable qubits and fixed coupler, to achieve high program success rate via software optimization of error mitigation.}
    \label{fig:tech}
\end{figure}

What is crosstalk? There is hardly a single precise noise model that captures all aspects of crosstalk, but rather, it is a combination of \emph{unwanted interactions between coupled qubits} on a quantum chip. This type of crosstalk noise prevails in many leading architectures, including trapped ion and superconducting systems \cite{craik2017high, ospelkaus2008trapped, neill2018blueprint, krantz2019quantum}. For superconducting transmon systems, two qubits interact with each other via \emph{resonance of qubit frequency}. Two main technology options for avoiding accidental resonance of qubits are: $i)$ to tune qubit frequencies apart using tunable qubits; $ii)$ to temporarily disable connections between qubits using tunable couplers. Fig.~\ref{fig:tech} illustrate the different design choices of leading QC architectures. Current IBM Q systems \cite{IBM} are built with fixed qubit frequency and fixed coupling, relying on a scheduler to avoid crosstalking gates \cite{murali2020software}; Google's architectures generally use tunable qubits with either fixed coupler \cite{barends2019diabatic} or tunable coupler \cite{arute2019quantum}. 

\begin{figure}[t]
    \centering
    \includegraphics[width=0.8\linewidth, trim=0 0cm 0 0]{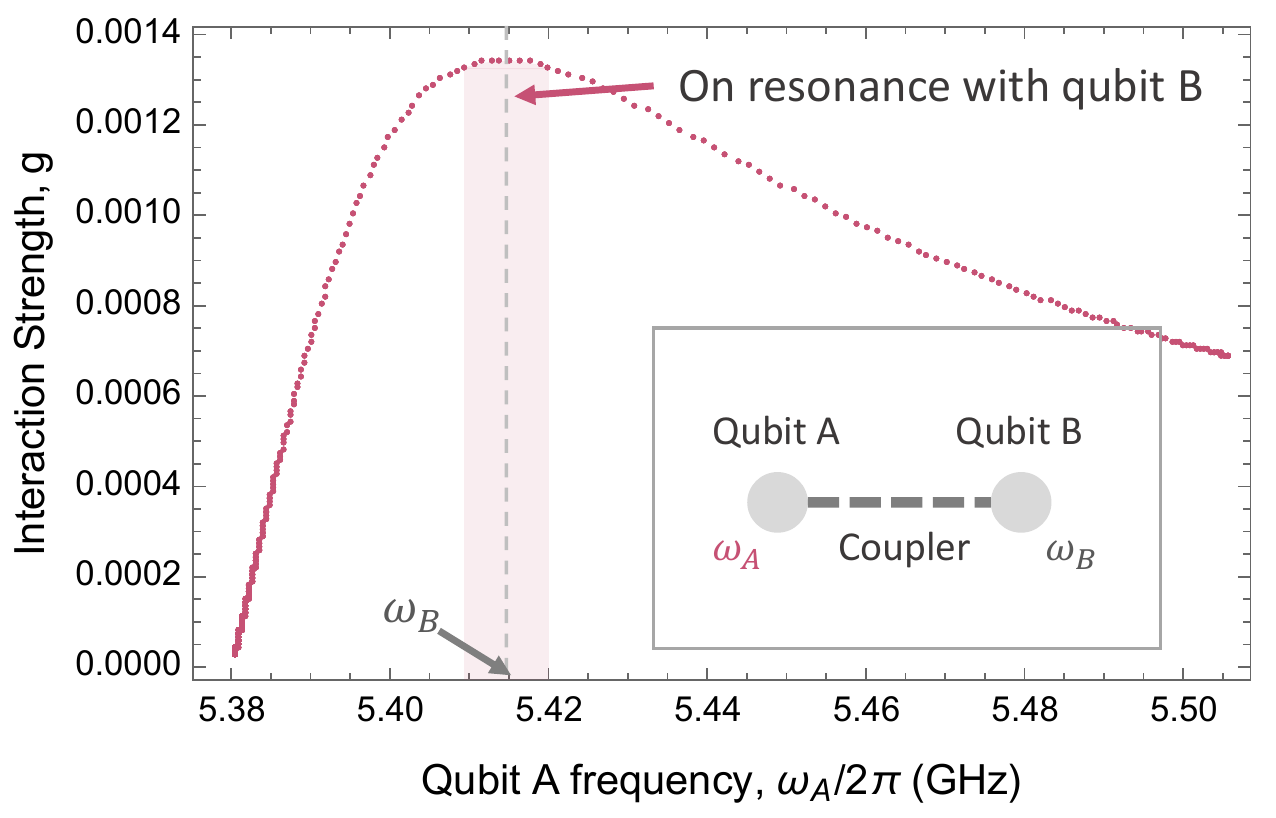}
    \caption{Interaction strength between two transmon qubits as we tune the frequency $\omega_A$ while holding $\omega_B$ constant. The strength peaks when two transmons are on resonance ($\omega_A=\omega_B$). Residual coupling remains when $\omega_A$ is close to $\omega_B$, and diminishes as $\omega_A$ is tuned far away from $\omega_B$. Inset: Schematics of two connected qubits.}
    \label{fig:coupling}
\end{figure}

Crosstalk noise is found to be highly dependent on the interaction strength between the qubits. For instance, Fig.~\ref{fig:coupling} shows the interaction between two connected (directly via a capacitor) frequency-tunable transmon qubits \cite{krantz2019quantum}. Unless the two qubit frequencies ($\omega_A$ and $\omega_B$) are tuned sufficiently apart, there remains some residual coupling between them, leading to unwanted crosstalk. 

When executing a quantum program, qubits are tuned dynamically to their assigned idle and interaction frequencies to perform single-qubit gates and two-qubit gates, respectively. As systems scale up and the frequency range becomes crowded, choosing frequencies for all qubits becomes increasingly challenging, necessitating compiler techniques for tuning frequencies systematically and scheduling instructions intelligently \cite{ding2020quantum}.

\begin{figure}
    \centering
    \includegraphics[width=0.7\linewidth, trim=0 0cm 0 0]{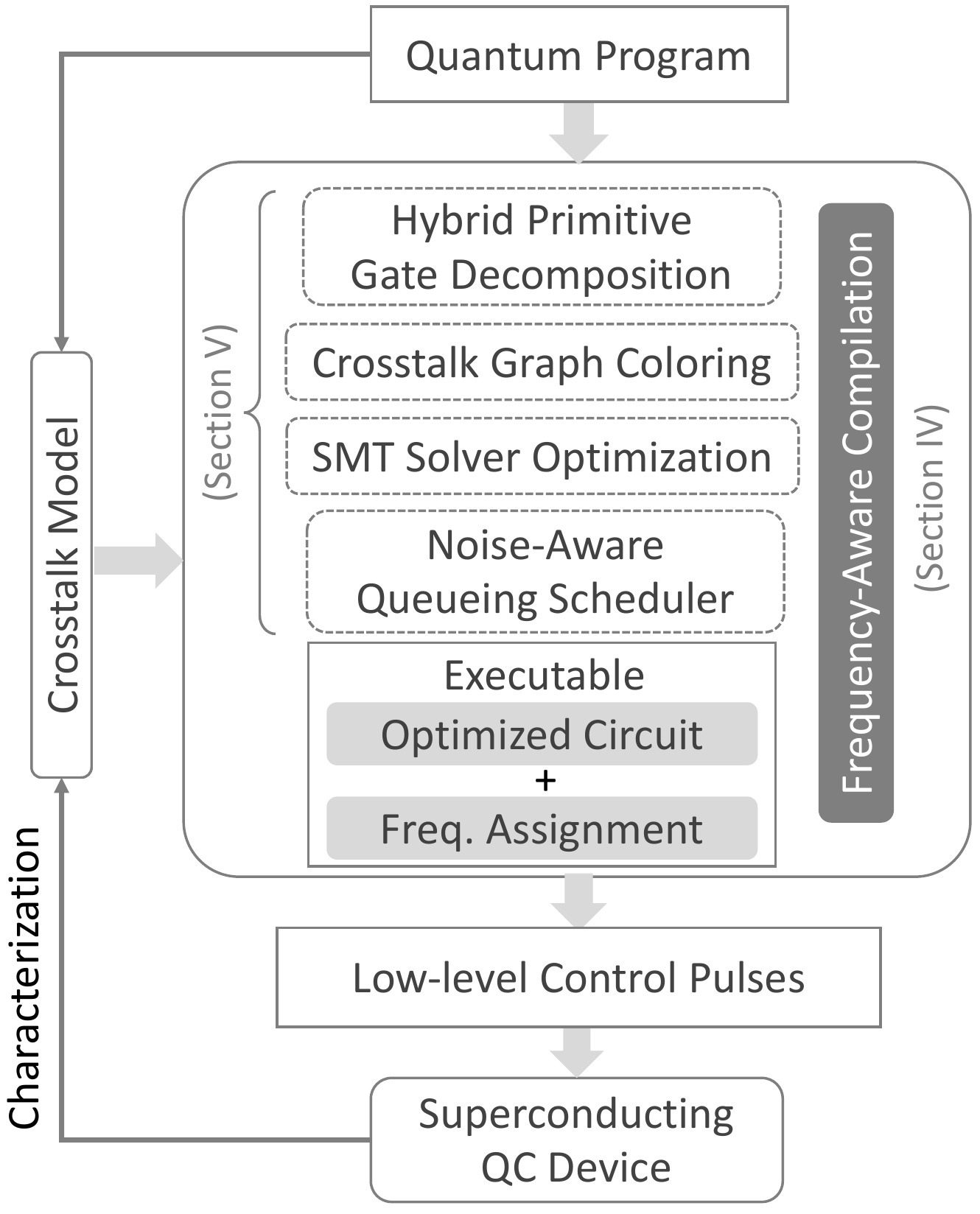}
    \caption{Flow of our crosstalk mitigation software for tunable superconducting QC systems. We develop a  frequency-aware compilation algorithm that systematically reduces crosstalk and decoherence.}
    \label{fig:flow}
\end{figure}

Fig.~\ref{fig:flow} is an overview of our approach. This work aims to provide means for understanding and mitigating the impact of crosstalk, from a software optimization perspective. Recent work by architects have demonstrated that software optimizations can lead to efficient noise mitigation, effectively providing the equivalent of months of hardware progress. For example, \cite{tannu2018case,li2019tackling,murali2019noise} show how to improve qubit utilization, and \cite{shi2019optimized, gokhale2020optimized} show how to optimize pulses to speedup gates. We demonstrate that quantum programs can be optimized to reduce the chance of crosstalk and decoherence by scheduling instructions at the right operational frequency and time step, preventing \emph{spectral} and \emph{temporal} collisions, respectively. To do so, we define a type of graph called the \emph{crosstalk graph}; our mitigation technique maps the frequency-aware compilation problem to the coloring of crosstalk graph. Furthermore, the diversity of gate decomposition gives us an extra degree of freedom in scheduling. In sum, our main contributions include:

\begin{itemize}
    \item An efficient compilation algorithm that mitigates the impact of crosstalk and decoherence via program-specific frequency tuning and instruction scheduling, making tunable-qubit, fixed-coupler systems a competitive, scalable design. 
    \item A systematic analysis of device tunability and sensitivity to provide insights on the advantages and disadvantages of different architectural designs, such as IBM's fixed-frequency systems and Google's tunable-coupler systems.
    \item Evaluations of our crosstalk mitigation algorithm on a variety of NISQ benchmarks including \texttt{BV} \cite{bernstein1997quantum}, \texttt{QAOA} \cite{farhi2014quantum}, \texttt{QGAN} \cite{lloyd2018quantum}, \texttt{ISING} \cite{barends2016digitized}, and \texttt{XEB} \cite{arute2019quantum} circuits.
\end{itemize}

The rest of the paper describes the details of our approach. Section~\ref{sec:bg} reviews the superconducting transmon architectures on which this work mainly focuses, and introduces gate operations and crosstalk noises on those architectures. Section~\ref{sec:related} compares the leading superconducting architectures and shows how the frequency-tunable qubit architecture is made competitive by our algorithm. Section~\ref{sec:frequency} is our proposed methodology for mitigating the frequency crowding problem, where we define the crosstalk graph and present our frequency tuning algorithm and circuit optimizations. Section~\ref{sec:approach} contains the implementation details of the proposed algorithm. Section~\ref{sec:eval} and Section~\ref{sec:results} evaluates our approach on a suite of NISQ algorithms. 
Finally, Section~\ref{sec:conclusion} discusses the implications and remaining issues.

%% file: background.tex
\section{Background}\label{sec:bg}
\subsection{Basics of Superconducting Qubits}

We start with a brief overview of superconducting qubits and how they are manipulated for computation. Transmon-like variety of superconducting qubits \cite{dicarlo2009demonstration, barends2014superconducting, kelly2015state, hutchings2017tunable} are among the most widely deployed quantum computer architectures \cite{arute2019quantum, reagor2018demonstration, krantz2019quantum}. The discussions in this work are centered around techniques for frequency-tunable transmons \cite{barends2014superconducting, kelly2015state, barends2016digitized, barends2013coherent}, but some general principles will be applicable to all types of superconducting architectures. 

\begin{figure}
    \centering
    \includegraphics[width=0.93\linewidth, trim=0 0cm 0 0]{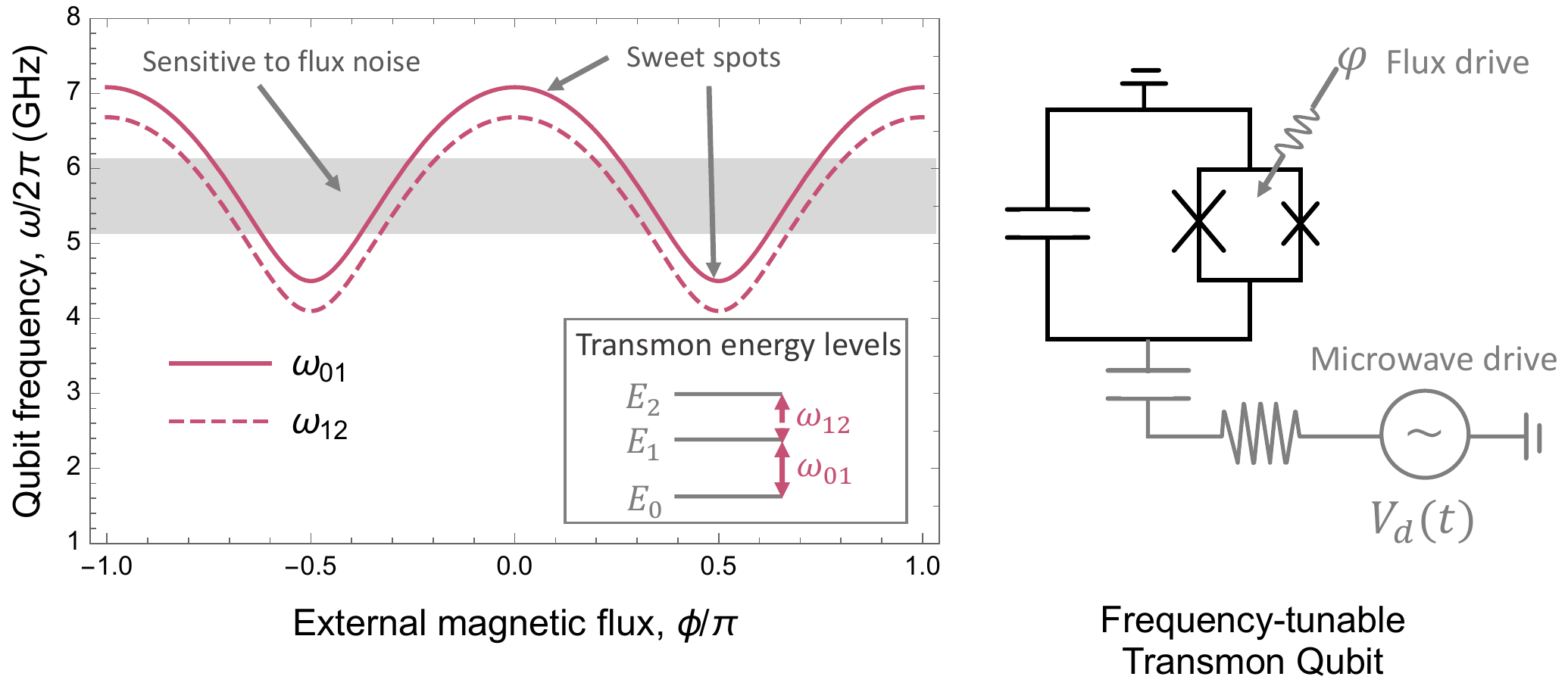}
    \caption{\textbf{Left:} Qubit frequencies as a function of external magnetic flux. The first three levels of the transmon, $\omega_{01}$ and $\omega_{12}$, are plotted. Shaded area is where the qubit is sensitive to flux noise. \textbf{Right:} Circuit diagram for a frequency-tunable (asymmetric) transmon qubit (highlighted in black), consisting of a capacitor and two asymmetric Josephson junctions. Highlighted in gray are two control lines: the external magnetic flux control $\varphi$ and microwave voltage drive line $V_d(t)$ for each transmon qubit.}
    \label{fig:transmon}
\end{figure}

A superconducting transmon \emph{quantum bit} (qubit), as shown in Fig.~\ref{fig:transmon}, is by design a multi-level quantum system made out of lithographically printed circuit elements, configured such that they exhibit atom-like energy spectra. The lowest two levels are used as the bit 0 and 1 for computation. The ground energy level represents the state $\ket{0} \equiv [1\;0]^T$, and the first excited energy level represents the state $\ket{1}\equiv [0\;1]^T$. Unlike a classical bit, a qubit can be in a linear combination of 0 and 1: $\ket{\psi} = \alpha\ket{0}+\beta\ket{1} = [\alpha\;\beta]^T$, where $\alpha, \beta$ are complex coefficients satisfying $|\alpha|^2 + |\beta|^2 = 1$. 

When a transmon gets accidentally excited to the second (or higher) energy level, e.g. $\ket{\psi} = \alpha\ket{0} + \beta\ket{1} + \gamma\ket{2}$, for $\gamma \neq 0$, we call this process ``leakage''. This can happen due to imprecision in quantum control. The energy gap between the ground state $\ket{0}$ and the first excited state $\ket{1}$ is known as the \emph{qubit frequency}, i.e. $\omega_{q} \equiv \omega_{01} = E_{01}/h$, where $h$ is the Planck's constant. Hence, we will sometimes use the terms energy and frequency interchangeably. More generally, $\omega_{01}$ is referred to as the (first-level) qubit frequency and $\omega_{02}$ is the second-level qubit frequency, defined as the gap between the ground state $\ket{0}$ and the second excited state $\ket{2}$. 
The frequency of a transmon qubit can be changed by applying external magnetic flux through the transmon loop, as shown in Fig.~\ref{fig:transmon}. 
In this case, there are two frequency sweet spots, i.e. frequency values that are relatively stable against flux noise \cite{krantz2019quantum}. As such, choosing operating frequencies around the sweet spots is desirable for tunable architectures.

\subsection{Operations and Noises}
In QC systems, computation is accomplished by applying a sequence of instructions/operations called \emph{quantum gates}, which take one quantum state to another through unitary transformations, i.e. $\ket{\psi} \rightarrow U\ket{\psi}$, where $U$ is a unitary matrix. These primitive transformations are implemented by driving the qubits via $i)$ microwave voltage signals, and $ii)$ local magnetic flux pulses. The control mechanism for each qubit is illustrated in Fig.~\ref{fig:transmon}. 

A \emph{quantum compiler} takes a quantum program written in a high-level programming language, performs a series of transformations and optimizations on the intermediate representations (IR) or quantum circuits, and finally outputs low-level control pulses for driving the qubits. At the end, results of the application are obtained by readout operations (called measurements) on the qubits, which collapse each qubit's quantum state to a classical bit $\ket{0}$ or $\ket{1}$.

\subsubsection{Single-qubit Gates and Decoherence Noise}



In superconducting transmon systems, single-qubit gates are implemented by driving the target qubit via: $i)$ a microwave drive line (feeding time-dependent voltage signals $V_d(t)$) through a capacitor connected to the qubit, and $ii)$ a flux drive line (with time-dependent magnetic flux pulses) \cite{krantz2019quantum}. For example, \texttt{Rx} and \texttt{Ry} rotation gates are implemented by sending microwave voltage signals in-phase (I) and out-of-phase (Q) through the drive line, respectively. 
Other single-qubit gates, such as Hadamard gate (\texttt{H}), can be accomplished by a combination of \texttt{Rx} and \texttt{Ry} gates. 

Qubits naturally decay due to perturbations from the environment. Such decay can happen in two ways: $i)$ \emph{$T_1$ relaxation} (i.e. spontaneous loss of energy causing decay from $\ket{1}$ to $\ket{0}$), and $ii)$ \emph{$T_2$ dephasing} (i.e. loss of relative quantum phase between $\ket{0}$ and $\ket{1}$). We can model both decays in a combined \emph{decoherence error}: $\epsilon_q(t) = (1-e^{-t/T_1})(1-e^{-t/T_2})$, where $t$ is time, and $T_1, T_2$ are constants characterizing the speed of the decays, for some qubit $q$.

\subsubsection{Two-qubit Gates and Crosstalk Noise}

\begin{figure}
    \centering
    \includegraphics[width=\linewidth, trim=0 0cm 0 0]{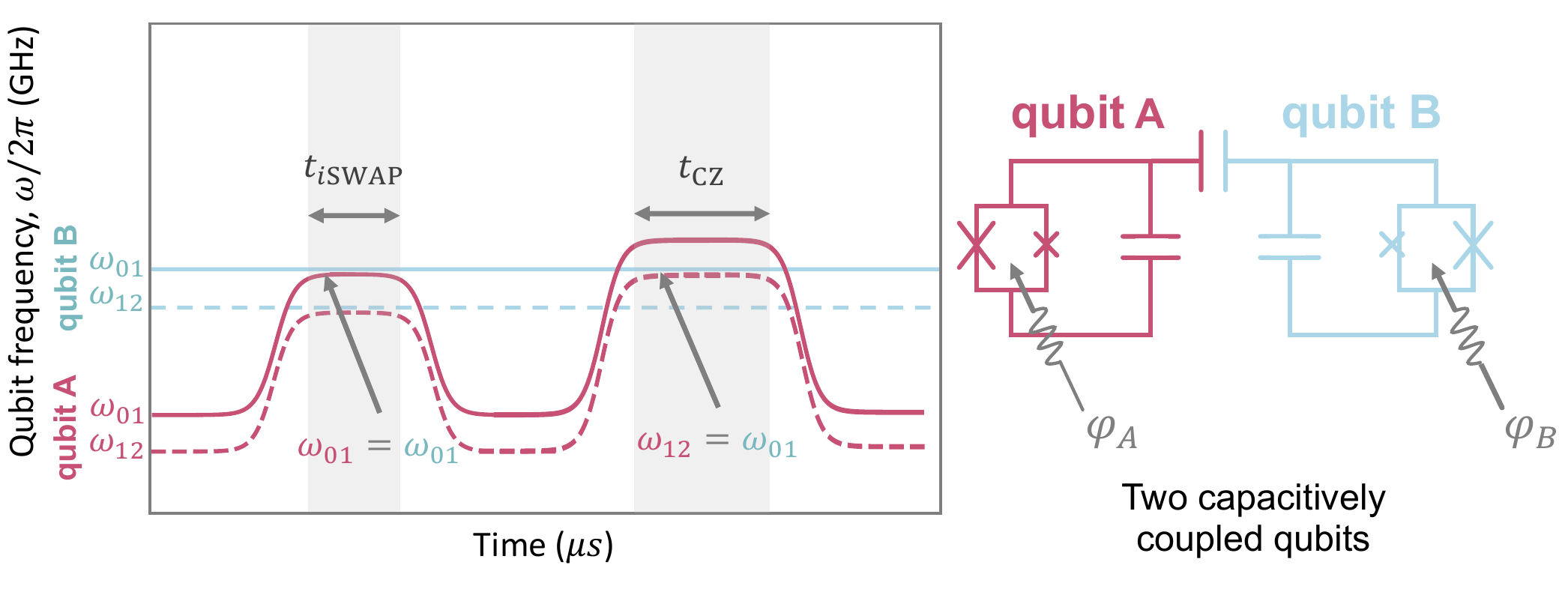}
    \caption{Two-qubit interactions for two capacitively coupled transmons. \textbf{Left:} 
    Two-qubit gates are implemented with resonance of qubit frequencies. Shown here are how qubit frequencies are tuned for $i$\texttt{SWAP} gate and \texttt{CZ} gate. \textbf{Right:} Circuit diagram of two capacitively coupled transmon qubits.}
    \label{fig:twotransmons}
\end{figure}

Two-qubit gates play important roles in quantum computation, as they implement entangling operations, that is, transformations of one qubit conditioned on the state of the other qubit \cite{caldwell2018parametrically}. Some commonly used two-qubit gates include \texttt{CNOT} (controlled-not) gate and \texttt{SWAP} gate. Despite their simple forms in the unitary matrix representations, these gates are not typically supported directly in the target architecture. For example, they need to be \emph{decomposed} into primitive gates, such as $i$\texttt{SWAP} gate and \texttt{CZ} (controlled-phase) gate, for tunable transmon architectures. The matrix forms for the $i$\texttt{SWAP} gate and the \texttt{CZ} gate are:

\begin{align*}
    i\texttt{SWAP} = \begin{bmatrix} 1 & 0 & 0 & 0 \\ 0 & 0 & -i & 0 \\ 0 & -i & 0 & 0 \\ 0 & 0 & 0 & 1 \end{bmatrix},\; \texttt{CZ} = \begin{bmatrix} 1 & 0 & 0 & 0 \\ 0 & 1 & 0 & 0 \\ 0 & 0 & 1 & 0 \\ 0 & 0 & 0 & -1 \end{bmatrix}.
\end{align*}

These gates are implemented by tuning the frequencies of the two interacting qubits to some desired operating point, denoted \emph{interaction frequencies}. Then, the qubits are held at that frequency for a duration of time $t$, depending on the interaction strength $g$ between the two qubits. Fig.~\ref{fig:twotransmons} depicts this process. Appendix~\ref{sec:tuning} explains the overhead of dynamically changing qubit frequencies.

In the most general sense, \emph{crosstalk} (i.e. unwanted interaction) happens when two qubits are accidentally tuned on (or close to) resonance. Fig.~\ref{fig:coupling} shows how interaction strength varies with closeness of frequencies, $\delta\omega = |\omega_A - \omega_B|$. Gate time $t$ is shorter when $g$ is higher (i.e. when $\delta\omega$ is small). Two-qubit gate error can therefore be modeled as a function of qubit frequencies and time: $\epsilon_g(\omega, t)$, for any gate $g$ (see Appendix~\ref{sec:append_gates} for details). For example, crosstalk can occur when a pair of two-qubit gates (on connected qubits simultaneously) happened to use very close interaction frequencies, as highlighted in Fig.~ \ref{fig:compiler_example}. Section~\ref{sec:frequency} illustrates in details how to understand and mitigate these types of crosstalk error.


%% file: related.tex
\section{Related Work}\label{sec:related}
A number of hardware features have been proposed to help mitigate crosstalk: $i)$ connectivity reduction, $ii)$ qubit frequency tuning, and $iii)$ coupler tuning. In addition to these hardware features, some software constraints are usually imposed to effectively reduce crosstalk; for example, certain operations may be prohibited to occur simultaneously.

\begin{figure*}
    \centering
    \includegraphics[width=0.93\textwidth, trim=0 0cm 0 0]{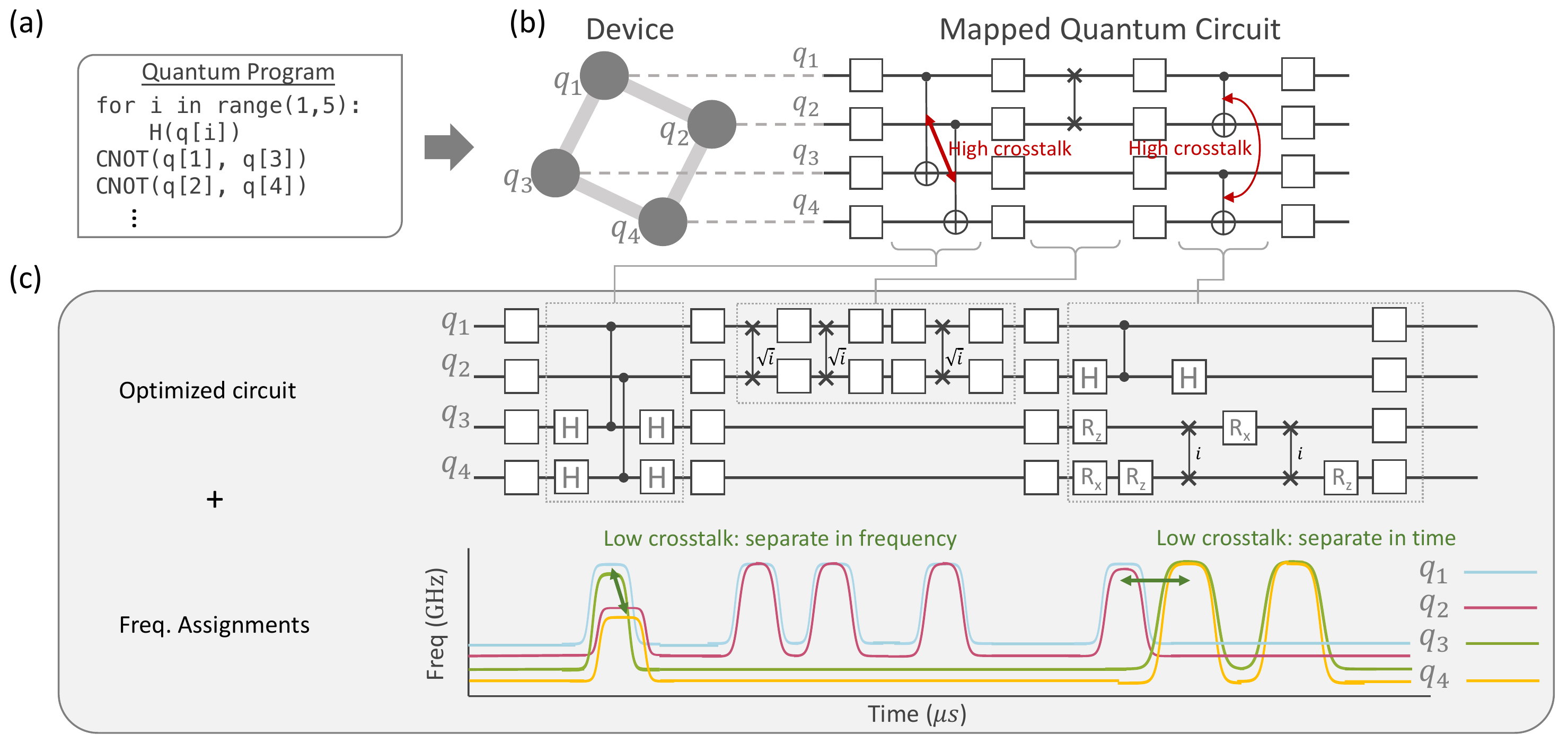}
    \caption{\textbf{(a)} An example quantum program on four qubits. \textbf{(b)} The quantum program is mapped to a QC system of $2\times 2$ qubits with nearest-neighbor connectivity. In a quantum circuit, qubits are lines; gates are applied to the qubits from left to right. Highlighted in red are the parallel quantum gates with high likelihood of crosstalk. \textbf{(c)} The optimized circuit and frequency assignment resulting from our compilation algorithm. Crosstalk is mitigated by avoiding spectral and temporal collisions in the those gates.}
    \label{fig:compiler_example}
\end{figure*}

\emph{Connectivity reduction} works by building devices with sparse connections between qubits, hence reducing the number of possible crosstalk channels. This greatly increases the circuit mapping and re-mapping overhead for executing a logical circuit, since many \texttt{SWAP} gates are needed. Moreover, this model necessitates an intelligent scheduler to serialize operations to avoid crosstalk \cite{murali2020software}. This strategy is commonly deployed for fixed-frequency transmon architectures, e.g. from IBM \cite{IBM}. Because of their non-tunable nature, these architectures have stringent constraints on the initial qubit frequency; a number of optimizers are proposed for this issue \cite{brink2018device, li2020towards}.

A second class of techniques rely on actively \emph{tuning qubit frequencies} to avoid crosstalk, featured in some prototypes \cite{hutchings2017tunable} and by Google \cite{barends2019diabatic}. Software can decide when to schedule an instruction and which frequency to operate the instruction at. In this class, \cite{versluis2017scalable} found a frequency assignment for the surface code circuit; \cite{helmer2009cavity} suggests a sudoku-style pattern of frequency assignment for cavity grid.

A third class builds not only frequency-tunable qubits but also \emph{tunable couplers} between qubits, termed ``gmon'' architectures \cite{chen2014qubit}. Without resorting to permanently reducing device connectivity in hardware, a different subset of connections are activated (via flux drives to the couplers) at different time steps. As such, a schedule for when to activate couplers is needed. After this work is submitted, \cite{klimov2020snake} outlines the frequency optimizer used in \cite{arute2019quantum}. Our results show comparable performance to \cite{klimov2020snake} but with simpler hardware (no tunable couplers). The control parameters used in \cite{klimov2020snake} are hard to predict, but in our evaluation, we include most of the leading noises, e.g., decoherence, sidebands resonance, leakages, flux noises, time overheads of flux tuning, etc.

Most previous studies on quantum program compilation \cite{shi2019optimized, gokhale2019partial} have largely targeted short program execution time (i.e. low circuit depth), and neglected the impact of gate errors such as crosstalk. Optimizations are performed at the gate level, typically involving strategic qubit mapping and instruction scheduling. Recent efforts \cite{murali2020software,li2020towards} are among the first to explore architects' role in mitigating crosstalk. 

Our work here shows that \emph{frequency-tunable architecture without connectivity reduction and without tunable couplers (but with our software crosstalk mitigation) is competitive against other architectures}. The frequency-tunable but untunable coupler architecture is an optimization sweet spot. On one end of the spectrum, fixed-frequency architectures have a relatively constrained space for software optimization. On the other end of the spectrum, requiring both qubit frequencies and couplers to be tunable introduces higher overhead in fabrication and higher control noises.


%% file: frequency.tex
\section{Systematic Crosstalk Mitigation}\label{sec:frequency}    

This work aims to demonstrate that \emph{systematic software optimizations} can dramatically mitigate crosstalk, utilizing a variety of microarchitecture tunability features. These features (such as different degree of tunability in qubits themselves and their couplers) allow the hardware to be \emph{dynamically configured} to avoid crosstalk as program executes. We propose frequency-aware software that reduces the chances of both decoherence and crosstalk, via strategic frequency tuning and instruction scheduling. 


\subsection{Understanding Crosstalk Constraints} \label{subsec:constraints}
Crosstalk mitigation is one of the major challenges in scaling up superconducting quantum architectures. Each qubit has a frequency $\omega_{q}^{01}$, as well as its associated higher-level excitation frequency $\omega_{q}^{12}$, which is slightly smaller than $\omega_{q}^{01}$. For qubit $A$ and qubit $B$ connected by a capacitor:
\begin{enumerate}[(i)]
    \item when qubits are non-interacting (i.e. during \texttt{Identity} or single-qubit gates), their \emph{idle frequencies} should have sufficient separation (e.g. $\omega_{A}^{01} \neq \omega_{B}^{01}$, $\omega_{A}^{01} \neq \omega_{B}^{12}$, and $\omega_{A}^{12} \neq \omega_{B}^{01}$);
    \item when implementing two-qubit gates, they should be placed on resonance at \emph{interaction frequency} (e.g. $\omega_{A}^{01} = \omega_{B}^{01}$ for \texttt{iSWAP} gate, and $\omega_{A}^{01} = \omega_{B}^{12}$ or $\omega_{A}^{12} = \omega_{B}^{01}$ for \texttt{CZ} gate).
\end{enumerate}


To avoid crosstalk, every pair of connected qubits must be fabricated or tuned to idle frequencies that satisfy the above constraints. However, each qubit can choose from a limited range \footnote{For example, in a typical  frequency-tunable transmon architecture, each qubit can be tuned to frequency around 5 GHz to 7 GHz \cite{arute2019quantum}.} of frequency spectrum. Furthermore, every two-qubit gate needs an interaction frequency far enough from those of its neighboring gates. This issue is termed \emph{frequency crowding}, because the frequencies grow increasingly crowded and the above constraints become harder to satisfy, as systems scale up and as programs use more parallelism. It is critical to determine the assignment of frequencies that minimizes unwanted crosstalk.

\subsection{Frequency Tuning and Instruction Scheduling}
To remedy this frequency crowding issue, we present a systematic scheme that dynamically tunes the device and schedules instructions according to input programs. Consider the toy program in Fig.~\ref{fig:compiler_example} as an example -- we found that \emph{a general recipe for avoiding crosstalk between two parallel gates is to create sufficient separation: $i)$ either in frequency, $ii)$ or in time.}  

In order to understand and mitigate the impact of crosstalk, we begin with two simple observations: $i)$ Every qubit (when not interacting with others) needs to pick a 0-1 excitation frequency sufficiently far apart from the 0-1 or 1-2 excitation frequencies of its neighbors. $ii)$ The extend of tunability is limited and there are few preferred operational frequencies for each qubit.
These two constraints are naturally in tension with each other. The key is to balance the two. 

To the best of our knowledge, this work is the first to study strategies for systematically tuning qubit frequencies in a program-aware fashion. 

Throughout the remainder of this paper, we explore crosstalk on a flux-tunable transmon architecture with 2-D mesh-like connectivity. Nonetheless, the input to our algorithm can be any arbitrary device topology; hence the crosstalk mitigation techniques we introduce here are applicable to all types of device connectivity, as showed quantitatively in Section~\ref{subsec:connectivity}.

\subsection{Resolving Frequency Crowding via Graph Coloring} \label{subsec:graph}
This section will focus on two types of graphs: $i$) the device connectivity graph, and $ii$) the crosstalk graph. For each of these two graphs, we will define formally and illustrate how coloring them can effectively reduce crowding of qubit frequencies. 

\subsubsection{Idle Frequencies and Connectivity Graph}
Qubit connectivity is an important characteristic of a quantum device, as it describes the pairs of qubits between which a two-qubit gate can be directly performed. For completeness, we revisit the definition of a connectivity graph: 
In a connectivity graph $G_c$, each vertex is a qubit, every edge is a coupling between the two qubits, e.g. a capacitor in the frequency-tunable transmon architecture. 

When the qubits are idle (i.e. not interacting with any other qubits), we want to avoid collision of frequencies for every pair of connected qubits. Therefore, we park the qubits at ``idle frequencies''. To avoid collisions in idle frequencies, it is equivalent to coloring the connectivity graph where no two end-points of an edge share the same color.  If a connectivity graph is colorable by $c$ colors, then we need only $c$ frequency values $\{\omega_{0},\omega_{1},\dotsc, \omega_{c-1}\}$ to keep idle qubits from interacting 
If the separation between the $c$ frequencies are large enough (i.e. any $|\omega_i-\omega_j|$ sufficiently larger than the anharmonicity), then the higher-energy excitation frequencies are also well separated from the other frequencies, reducing interactions through the leakage channel as well. This strategy works well for simple connectivity graphs like the 2-D mesh, because the 2-D mesh is bipartite and thus 2-colorable. We also test the general applicability of our algorithm on different choices of device connectivity.

\begin{figure*}
    \centering
    \includegraphics[width=0.8\linewidth, trim=0 0cm 0.5cm 0]{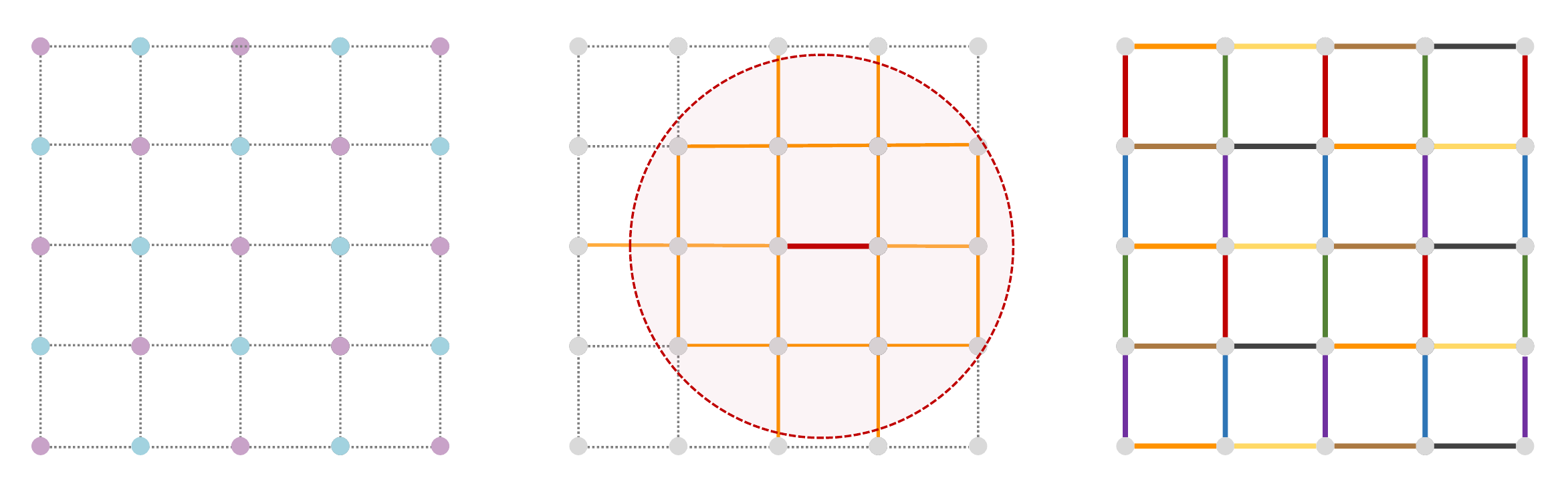}
    \caption{\textbf{Left:} the connectivity graph for a $5 \times 5$ mesh of qubits; 2 colors (highlighted in blue and purple) are needed to color the nodes of the graph. The colors map to idle frequencies of the qubits. \textbf{Center:} when the two qubits at the center choose an interaction frequency (highlighted in red) all qubits within the crosstalk range must be tuned off resonance from this interaction frequency. \textbf{Right:} A non-crosstalking edge coloring of the 2-D mesh, resulting from coloring the crosstalk graph. 8 colors are required to avoid crosstalk among maximum simultaneous operations. Notably, fewer colors will suffice for program-specific compilation that utilizes circuit slicing and subgraph coloring.}
    \label{fig:gridcoloring}
\end{figure*}

\subsubsection{Interaction Frequencies and Crosstalk Graph}
Two-qubit gates are implemented by bringing the two qubits on resonance at some ``interaction frequency''. Any other qubits nearby should be tuned off-resonance from that frequency to avoid unwanted interactions. We define the crosstalk graph to exactly match this constraint.
The crosstalk graph $G_x$ of a connectivity graph $G_c$ represent the potential crosstalk that could happen between qubits, which must be addressed by frequency tuning. Here we describe how to construct the crosstalk graph $G_x$:
\begin{enumerate}[(i)]
    \item Derive the line graph\footnote{A line graph of a graph $G$ maps each edge in $G$ to a vertex, and two vertices are connected if the two edges in $G$ share a same vertex. \cite{harary1960some}} $G_L$ of the connectivity graph $G_c$.
    \item Connect two vertices in $G_L$ if the corresponding two edges in $G_c$ is distance\footnote{Distance between two edges equals the length of the shortest path that connects the two edges.} one apart.
\end{enumerate}

To elucidate the structures behind the crosstalk graph, we use a $5\times 5$ quantum chip as an example. Consider the middle edge highlighted in red in the center panel of Fig.~\ref{fig:gridcoloring}. Every orange edge either shares a common vertex with the red edge or is connected to the red edge by a third edge. Thus in the crosstalk graph, the vertex corresponding to the red edge in $G_c$ is connected with the vertices corresponding to all orange edges.  If we tune the qubits on the red edge to an interaction frequency $\omega_{int}$, then during the gate time, none of the orange edges should share that frequency. 

Although quite dense (see Fig.~\ref{fig:augmented_line_graph}), the crosstalk graph for a 2-D mesh can be colored by 8 colors as shown at the right of Fig.~\ref{fig:gridcoloring}. This coloring is general for any $N\times N$ 2-D mesh, and 8 is the minimum number of colors needed. See Appendix~\ref{sec:example_color} for an example of idle and interaction frequencies resulting from coloring crosstalk graph.

We report an important observation here: for a device with 2-D mesh connectivity, crosstalk due to frequency crowding is \emph{mostly localized}. In other words, the frequency space does not become more and more crowded as we increase the size of the mesh. To understand how localized is it, we extend our discussion on nearest-neighbor crosstalk to \textit{next}-neighbor crosstalk. 

\subsubsection{Generalization to Higher Distance}
So far, we have been discussing crosstalk between directly coupled qubits (i.e. nearest-neighbor crosstalk). One could imagine the residual coupling between a qubit and its next-neighbor could result in crosstalk as well. We introduce a generalization to the crosstalk graph to higher distance $d$, denoted as $G_x^{(d)}$:
The distance-$d$ crosstalk graph $G_x^{(d)}$ of a connectivity graph $G_c$ has a vertex for each edge in $G_c$, and two vertices are connected if the two edges in $G_c$ share a common vertex or are connected by a path of length $d$.


%% file: approach.tex
\section{Our Approach}\label{sec:approach}

 
\subsection{Frequency-Aware Compilation: Overview}
Now we illustrate the key steps in our crosstalk mitigation algorithm -- the inputs to the algorithm include device characteristics (e.g. qubit number, connectivity, transmon tunability), program characteristics (e.g. a scheduled quantum circuit), and optimization level (e.g. crosstalk distance). 

Finding optimal (idle and interaction) frequency configurations based on device and program characteristics is a high-dimensional optimization problem; \emph{we break the problem into multiple scalable sub-problems}. As shown in Fig.~\ref{fig:flow}, we begin by constructing a crosstalk graph for the input device. Next, the input program is decomposed into primitive gates and sliced into layers (time steps). Then, we produce \emph{a feasible coloring of an active subgraph} of the crosstalk graph for each layer of the circuit. From the colors, we thereafter map to the idle and interaction frequencies via a Satisfiability Modulo Theory (SMT) solver \cite{bjorner2015nuz, de2008z3}. Lastly, we produce a feasible schedule of the program (i.e. gate instructions and qubit frequencies for each time step), throttling parallelism if necessary. Algorithm~\ref{algo:main} is the main algorithm outlining this process. Specifically, line 10-16 is the queueing schedule in Section~\ref{subsec:scheduler}; line 17-19 is the coloring step in Section~\ref{subsec:coloring}; line 20-22 corresponds to the SMT solver optimization in \ref{subsec:smt}.

\begin{algorithm}[t]
\small
 \caption{Frequency-Aware Compilation}
 \label{algo:main}
\begin{algorithmic}[1]
\State $d \gets$ crosstalk distance parameter
\State ${G}_c \gets$ connectivity graph of the device ${D}$
\State ${G} \gets$ gen$\_$crosstalk$\_$graph(${D}$, d)
\State ${C}_{c} \gets$ coloring(${G}_c$)
\State $\Omega_c \gets$ colors in ${C}_{c}$ are mapped to parking frequencies
\State ${P} \gets$ decompose input program ${P}$ into primitive gates

\State $S \gets$ first layer (time step) of program $P$
\State $Q \gets \varnothing$
\While {${S}$ non empty}
 \State $I \gets \varnothing$
 \State $S \gets \text{sort S by criticality}$
 \For{gate in $S$}
  \If{not noise$\_$conflict(gate, I)}
    \State ${I} \gets {I} \cup \{gate\}$
  \EndIf
 \EndFor
  \State $E \gets \text{collect relevant two-qubit gates in I}$
  \State ${H} \gets \text{subgraph}({G}, E)$
  \State ${C} \gets \text{coloring}({H})$ 
  \State $\Omega \gets \text{smt\_find}({C})$
  \State $S \gets (S \setminus I) \cup \{\text{next layer of } P\}$
 \State $F \gets $ qubit frequencies for this cycle based on $\Omega_c$ and $\Omega$
 \State $Q \gets Q \cup \{(I, F)\}$
\EndWhile
\State \Return $Q$
 \end{algorithmic}
\end{algorithm}

\subsection{Optimization Details}\label{sec:implement}
This section is dedicated to explaining the key ingredients of the algorithm in greater detail. Through a series of optimizations, our frequency-aware compilation algorithm drastically reduces the chance of crosstalk and scales favorably with systems sizes, making it a viable long-term solution to frequency tuning for superconducting qubits. 

\subsubsection{Crosstalk Graph Construction}
In Section~\ref{subsec:graph}, we outlined how the crosstalk graph is constructed; the steps are made rigorous in the following Algorithm~\ref{algo:graph}. By abstracting all possible crosstalk channels between pairs of qubits as graph theoretical objects, we are now equipped to quantitatively analyze and systematically mitigate crosstalk errors due to frequency crowding.

\begin{algorithm}[t]
\small
 \caption{gen\_crosstalk\_graph}
 \label{algo:graph}
\begin{algorithmic}[1]
\State ${G}_c \gets$ connectivity graph of the device ${D}$
\State ${G} \gets$ networkx.line$\_$graph(${G}_c$)
\State ${S}  \gets \varnothing$
 \For{pair of nodes $(e_1,e_2)$ in ${G}_\ell$}
  \State $(u_1, v_1) \gets \text{pair of qubits for } e_1$
  \State $(u_2, v_2) \gets \text{pair of qubits for } e_2$ 
  \State $\text{cond} \gets dist(u_1, u_2) \leq d \text{ or } dist(u_1, v_2) \leq d$
  \State $\text{cond} \gets \text{cond} \text{ or } dist(v_1, u_2) \leq d \text{ or } dist(v_1, v_2) \leq d$
  \If{cond}
    \State ${S} \gets {S} \cup \{(e_1, e_2)\}$
  \EndIf
 \EndFor
 \State ${G}$.add$\_$edges$\_$from(${S}$)
 \State \Return $G$
 \end{algorithmic}
\end{algorithm}

\subsubsection{Circuit Slicing and Subgraph Coloring}\label{subsec:coloring} 
One of the major advantages of our approach is in producing a \emph{dynamic} frequency assignment tailored for each input program. This wins over a static (program independent) frequency assignment because frequencies are substantially less crowded when only considering a subset of couplings between qubits that are ``active'' for a given time step. Here active couplings refers to only those pairs of qubits currently involved in two-qubit gates. 

We identify the active subgraph $H$ of the crosstalk graph $G$, by profiling the two-qubit gates in one time step. The (vertex) coloring of $H$, denoted as $C$, is an assignment of labels (called colors) for the vertices of $H$ such that no two adjacent vertices share the same color, while minimizing the number of colors in total. Graph coloring is known to be an NP-complete problem; section~\ref{subsec:scalability} shows how we maintained efficiency. In our optimization, we apply a polynomial-time greedy approximation, the Welsh-Powell algorithm \cite{welsh1967upper}, to color the active subgraph.

As a result, a feasible coloring of $H$ yields a set of non-colliding interaction frequencies for the two-qubit gates. Qubits that undergo \texttt{Identity} or single-qubit gates are parked at idle frequencies, determined by coloring the device connectivity graph. In the next section, we describe how to map from a coloring to a frequency assignment via a SMT solver. 

\subsubsection{SMT Solver Optimization}\label{subsec:smt}

The mapping from colors $C$ to frequencies $\Omega$ is reduced to a constrained optimization problem. The objective is to assign $|C|$ frequencies within some range $[\omega_{lo}, \omega_{hi}]$, satisfying the crosstalk constraints in Section~\ref{subsec:constraints}. We use a SMT solver to find a feasible solution with the following constraints.
\begin{align}
    \forall c \in C,& \omega_{lo} \leq x_c \leq  \omega_{hi},\\
    \forall x_{c_i},& x_{c_j}, |x_{c_i}-x_{c_j}|\geq\delta,\\
    & |x_{c_i}+\alpha-x_{c_j}|\geq\delta,
\end{align}
where $\alpha$ is the anharmonicity, and $\delta$ is a threshold. Then, \texttt{smt\_find} uses a simple binary search to find the maximum threshold $\delta$, for which a feasible solution exists. We ensure the efficiency of the procedure by keeping $|C|$ small. 

Once the optimal solution is found, a one-to-one mapping from $C$ to $\Omega$ is enforced by a total ordering, motivated by the fact that higher interaction frequency value would yield faster gate time, i.e., $t_{gate} \sim 1/\omega$ \cite{krantz2019quantum}. In particular, let us denote $n(c)$ as the number of times $c$ appear in $C$ and $\omega(c)$ as the frequency value to which $c$ maps. We dictate that, for any $c_i, c_j \in C$, if $n(c_i) \geq n(c_j)$ then $\omega(c_i) \geq \omega(c_j)$. The following section details how the frequency ranges are determined. 

\subsubsection{Frequency Partitioning}\label{subsec:partition}
We partition the range of tunable frequency spectrum into three regions: interaction region, exclusion region, and parking region. Similar partitioning strategies has been studied for surface code error correction circuits \cite{versluis2017scalable}. This allows us to decouple the idle frequency assignment from that of the interaction frequency. For a realistic frequency-tunable transmon, the tunable range is typically just a few GHz. So a reasonable design would use a partition with 1 GHz interaction region, 0.5 GHz exclusion region, and 1 GHz parking region. By this design, no frequency is assigned in the exclusion region (which are most sensitive to flux noise), preventing idle qubits from interacting with iswap/cphase qubits.

The interaction frequencies are determined using the coloring $C$ for $H$. This is a two-step process. First, each coupling in $H$ (that is a pair of qubits performing a two-qubit gate) gets assigned a color $c \in C$ corresponds to an interaction frequency. Second, qubits that appear in its complement $G\setminus H$ remain in their parking frequencies.

\subsubsection{Hybrid Circuit Decomposition}
\begin{figure}
    \centering
    \includegraphics[width=0.8\linewidth, trim=0 0cm 0 0]{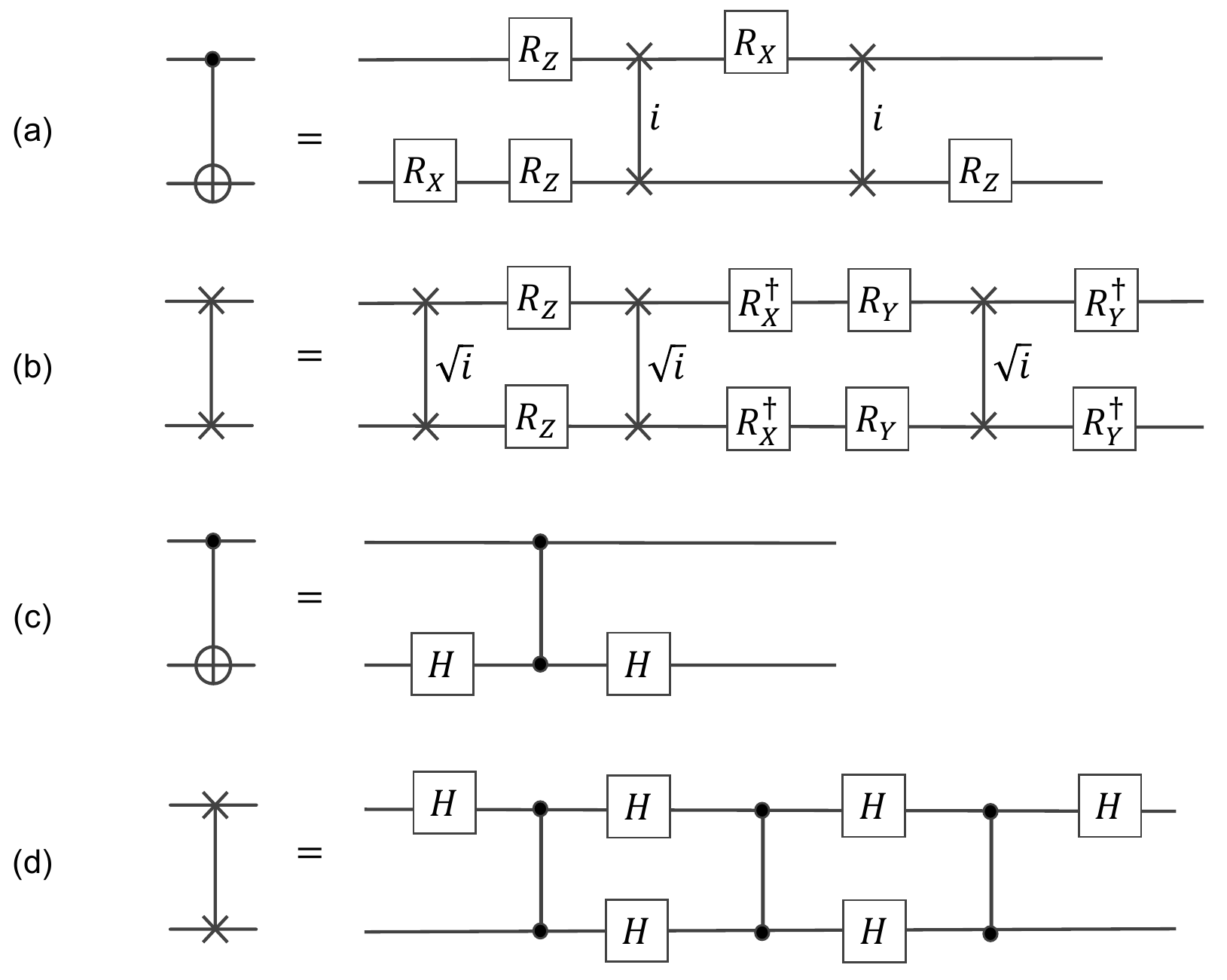}
    \caption{\textbf{(a)}: The \texttt{CNOT} gate, decomposed with $i$\texttt{SWAP}. \textbf{(b)}: The \texttt{SWAP} gate, decomposed with $\sqrt{i\texttt{SWAP}}$. \textbf{(c)}: The \texttt{CNOT} gate, decomposed with \texttt{CZ}. \textbf{(d)}: The \texttt{SWAP} gate, decomposed with \texttt{CZ}.}
    \label{fig:circuitdecomp}
\end{figure}
To implement a two-qubit gate that is not directly supported by the frequency-tunable transmon architecture, we need to decompose it into a series of native gates. Two commonly used two-qubit gates in quantum programs are the \texttt{CNOT} gate and the \texttt{SWAP} gate, because they implement relatively simple Boolean logic. Fig.~\ref{fig:circuitdecomp} shows that they can be decomposed into \texttt{iSWAP} (or $\sqrt{i\texttt{SWAP}}$) and \texttt{CZ} gates.

The strategy for circuit decomposition can affect performance. Compared to decomposing all the two-qubit gates in a circuit with one type of native gates, hybrid strategies can help achieve better fidelity. A simple hybrid strategy is to decompose \texttt{CNOT} gates with \texttt{CZ}, and \texttt{SWAP} gates with $\sqrt{i\texttt{SWAP}}$. As depicted in Fig.~\ref{fig:circuitdecomp}, this strategy is advantageous because \texttt{CNOT} (\texttt{SWAP}) is cheaper to implement with \texttt{CZ} ($\sqrt{i\texttt{SWAP}}$) gates than with $\sqrt{i\texttt{SWAP}}$ (\texttt{CZ}) gates.

\subsubsection{Noise-Aware Queueing Scheduler}\label{subsec:scheduler}
Finally, parallelism is another crucial concern in our algorithm -- on one hand, parallelism helps shorten the circuit execution time, reducing chances of decoherence; on the other hand, it crowds the interaction frequency range, increasing chances of crosstalk. Our noise-aware queueing scheduler finds a sweet spot by strategically serializing gates that are likely to cause crosstalk. In algorithm~\ref{algo:main} (line 9-16), gates are delayed based on their criticality and potential noise conflicts. Criticality of a gate is its position along the program critical path, calculated by profiling the input program during circuit slicing on line 7. Function \texttt{noise$\_$conflict} predicts potential crosstalk: when scheduling $g$ (e.g. \texttt{CNOT(q1,q2)}), if too many of its neighbors in the crosstalk graph are already in $I$, then their interaction frequencies are likely very close, so we postpone $g$ for the next time step. Serialization is done conservatively while maintaining minimal impact on the critical path length of the program (that is the circuit depth). This \emph{greedy} scheduling approach is shown to be effective in balancing crosstalk and decoherence.


%% file: evaluation.tex
\section{Evaluation}\label{sec:eval}

\begin{table}[t]
    \caption{List of algorithms used in our evaluation}
    \footnotesize
    \centering
    \begin{tabular}{c|M{5.4cm}}
    \hline\hline
        Algorithms & Microarch. Features  \\[0.1em]\hline
        Baseline N & Tunable transmon, fixed coupler, Qiskit \cite{Qiskit} scheduler \\\hline

        Baseline G & Tunable transmon, tunable coupler, tiling scheduler  \\\hline
        Baseline U & Tunable transmon (with single interaction frequency), fixed coupler, serial scheduler \\\hline
        Baseline S & Tunable transmon, fixed coupler, crosstalk-aware scheduler   \\\hline
        ColorDynamic & Tunable transmon, fixed coupler, crosstalk-aware scheduler  \\\hline
    \hline
    \end{tabular}
    \label{tab:baselines}
\end{table}

\subsection{Tuning and Scheduling Baselines}
We test the performance of our frequency-aware compilation algorithm (i.e. \emph{ColorDynamic}) in comparison to four baselines, \emph{Baseline N} (naive), \emph{Baseline G} (gmon), \emph{Baseline U} (uniform), and \emph{Baseline S} (static), shown in Table~\ref{tab:baselines}; they represent strategies of frequency tuning and instruction scheduling from leading industry architectures. 

\vspace{0.2cm}\noindent\textbf{Baseline N: Naive Compilation.} A conventional crosstalk-unaware compilation algorithm. Qubits are assigned with separated idle and interaction frequencies.

\vspace{0.2cm}\noindent\textbf{Baseline G: Gmon with Tunable Coupler. } This baseline has advanced hardware requirements to activate couplers -- the ``gmon'' architecture, implemented in Google's recent Sycamore quantum architectures \cite{arute2019quantum}, takes advantage of both tunable qubit and tunable coupling features to mitigate crosstalk. On the flip side, the flux-tunable coupler would incur fabrication overheads, and introduce extra sensitivity to flux noise. We reconstruct and evaluate a gmon-like architecture where the couplers are activated following the same pattern used for Sycamore, and idle and interaction frequencies match exactly the reported values in \cite{arute2019quantum}.

\vspace{0.2cm}\noindent\textbf{Baseline U: Uniform Frequency with Serialization. } This baseline relies on serialization to avoid crosstalk, similar to \cite{IBM, murali2020software}. All two-qubit gates share one common interaction frequency $\omega_{int}$, demonstrating the impact of serialization.

\vspace{0.2cm}\noindent\textbf{Baseline S: Static Frequency-Aware Compilation. } Baseline S optimizes the idle and interaction frequencies independent of input programs, producing a static set of optimized values. Most crosstalk-aware optimizers perform this type of static optimization \cite{versluis2017scalable,arute2019quantum}.

\vspace{0.2cm}\noindent\textbf{ColorDynamic: Program-specific Frequency-Aware Compilation. } This is the pinnacle of our work. Instead of finding a static interaction frequency solution for all programs, ColorDynamic returns optimized frequencies for each time step of a program. It combines all optimizations in Algorithm~\ref{algo:main}, including circuit slicing, strategical decomposition and serialization, graph coloring, and SMT solvers.

\subsection{Benchmarks}
We study the performance of our algorithm through a variety of NISQ benchmarks, shown in Table~\ref{tab:benchmarks}. These benchmarks are among the best known applications for near-term quantum machines. We also include circuits for benchmarking simultaneous quantum gates to demonstrate the impact of crosstalk on the fidelity of those gates \cite{arute2019quantum}. 

In our evaluation, we vary number of qubits $n=4,9,16,25$. These circuits are of most interest, because the range of crosstalk is typically localized, as shown in Fig.~\ref{fig:gridcoloring}.  

\begin{table}[htb]
    \caption{List of benchmarks used in our evaluation}
    \footnotesize
    \centering
    \begin{tabular}{c|M{6.1cm}}
    \hline\hline
        Benchmarks & Descriptions \\\hline
        \texttt{BV(n)} & Bernstein-Varzirani (BV) algorithm on $n$ qubits \cite{bernstein1997quantum} \\\hline
        \texttt{QAOA(n)} & Quantum Approximate Optimization Algorithm (QAOA) \cite{farhi2014quantum} for MAX-CUT on an Erdos-Renyi random graph with $n$ vertices\\\hline
        \texttt{ISING(n)} & Linear Ising model simulation of spin chain of length $n$ \cite{barends2016digitized} \\\hline
        \texttt{QGAN(n)} & Quantum Generative Adversarial Network (QGAN) with training data of dimentsion $2^n$ \cite{lloyd2018quantum}\\\hline
        \texttt{XEB(n, p)} & Cross entropy benchmarking circuit for calibrating two-qubit gates on $n$ qubits with $p$ cycles \cite{arute2019quantum} \\
    \hline\hline
    \end{tabular}
    \label{tab:benchmarks}
\end{table}

\subsection{Experimental Setup}
\vspace{0.2cm}\noindent\textbf{Software implementation: } Our compilation algorithms are implemented in \texttt{Python} 3.7, interfacing the IBM \texttt{Qiskit} software library \cite{Qiskit}. The graph coloring optimization uses \texttt{greedy\_coloring} in \texttt{NetworkX} library \cite{hagberg2008exploring}, and the SMT optimization uses \texttt{Z3 solver} \cite{de2008z3} through the \texttt{Z3py} APIs. All compilation experiments use Intel E5-2680v4 (2.4GHz, 64GB RAM). 

\vspace{0.2cm}\noindent\textbf{Architectural features: } We consider a 2D grid of $N \times N$ asymmetric frequency-tunable transmons, each having maximum frequencies $\omega_q$ (in GHz) sampled from Gaussian distribution: $\Omega \sim \mathcal{N}(\omega, 0.1)$, with nearly constant aharmonicity $\alpha/2\pi = (\omega_{12}-\omega_{01})/2\pi \approx 200$ MHz, to account for realistic variation in fabrication and initial detuning. Any pair of nearest-neighbor qubits are directly connected with a capacitor; the coupling strength $g$ depends on the frequencies of the qubits, which is typically around $g/2\pi \approx 30$ MHz. For gmon-like experiments, qubits are connected by flux-tunable couplers, each with its own independent external magnetic flux control. These parameters are set to realistic values in line with experimental data from the literature \cite{kjaergaard2020quantum}.


\vspace{0.2cm}\noindent\textbf{Metrics: } For our compilation experiments, we need to efficiently compute the program success rate -- we define a heuristic for efficiently estimating the \emph{worst case} success rate of a program under crosstalk and decoherence noises. 
\begin{align}
    P_{success} = \Pi_{g \in G}(1-\epsilon_g) \cdot \Pi_{q \in Q} (1-\epsilon_q) \label{eq:psuccess}
\end{align}
where $\epsilon_g$ is the crosstalk gate error, and $\epsilon_q$ is qubit decoherence error. Details on $\epsilon_g$ can be found in Appendix~\ref{sec:append_gates}, equation~\ref{eq:eg}; $\epsilon_q$ is captured by modeling $T_1$ and $T_2$ during idle or gate time, as studied in \cite{kjaergaard2020quantum}. A similar metric to $P_{success}$ is used in \cite{zlokapa2020boundaries, arute2019quantum}.

Besides being efficiently computable, this heuristic has useful operational significance -- we can understand and mitigate the worst-case impact of crosstalk and decoherence on the systems during compile-time or run-time of quantum programs. Of course, to gain full knowledge of the crosstalk and decoherence errors, we need full noisy circuit simulation, which quickly becomes intractable as circuit size grows beyond tens of qubits. Hence, we validate the heuristic estimator on small-scale circuits, for which noisy circuit simulation is possible. 


\begin{figure*}
    \centering
    \includegraphics[width=0.97\textwidth, trim=0 0.2cm 0 0cm]{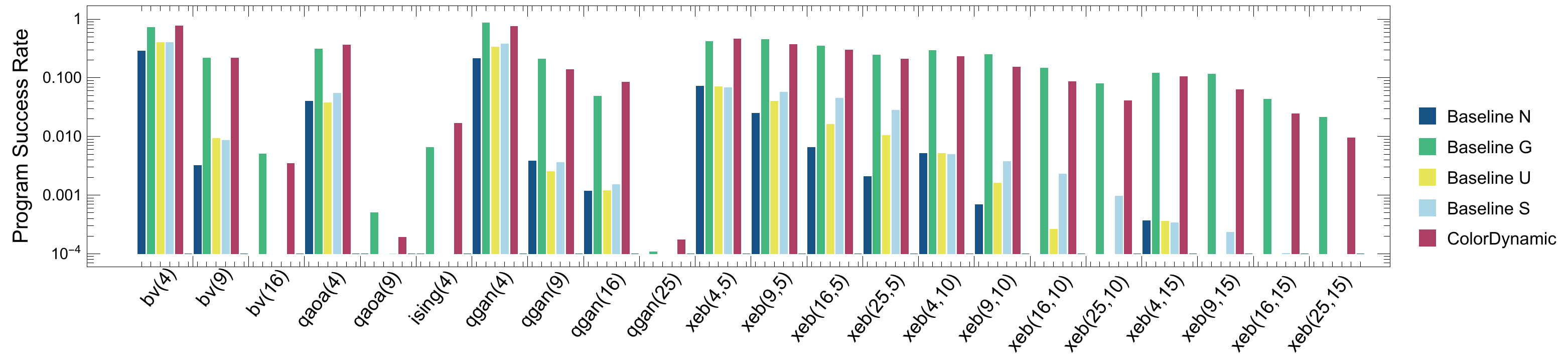}
    \caption{Log-scale worst-case program success rates using crosstalk-mitigation algorithms, estimated by heuristics. Higher success rate is better. Across the benchmarks, ColorDynamic performs consistently well compared to other algorithms. In particular, it matches the crosstalk resilience of baseline G (with tunable-qubit, tunable coupler), but on fixed-coupler hardware which is more robust to external noise. Results for qaoa(16) and ising(16) are omitted due to high circuit depth and qubit decoherence. }
    \label{fig:success_rates}
\end{figure*}
\begin{figure*}
    \centering
    \includegraphics[width=0.9\textwidth, trim=0 0.2cm 0 0]{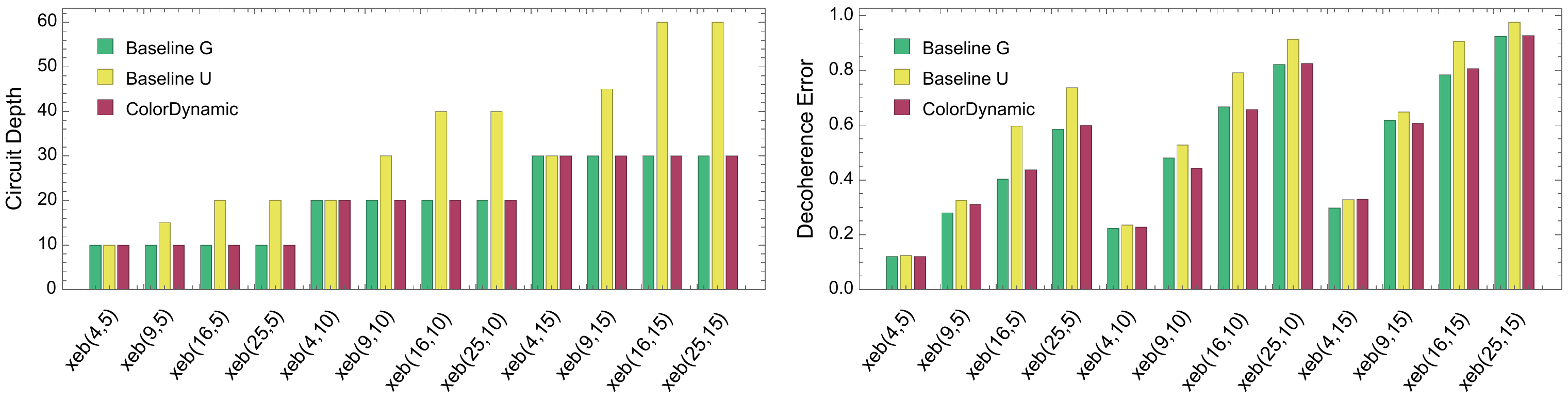}
    \caption{\textbf{Left:} Circuit depth resulting from crosstalk-mitigation algorithms. Across the benchmarks, ColorDynamic avoids crosstalk without incurring significant serialization. \textbf{Right:} Decoherence errors resulting from crosstalk-mitigation algorithms. Lower is better.}
    \label{fig:depths}
\end{figure*}

%% file: results.tex
\section{Results} \label{sec:results}

\subsection{Program Success Rate}
\begin{figure}[t]
    \centering
    \includegraphics[width=0.9\linewidth, trim=0 0cm 0 0]{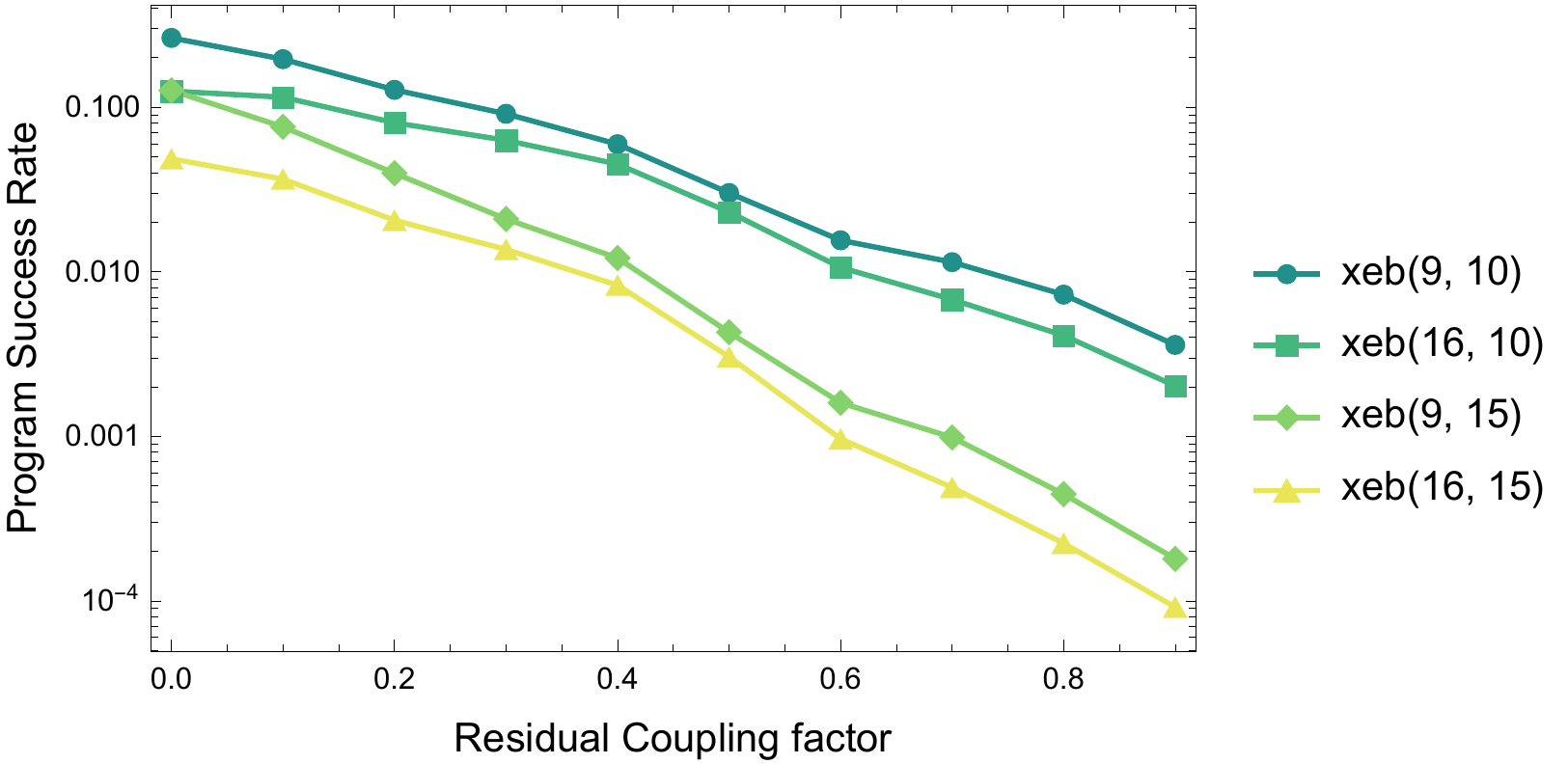}
    \caption{Log-scale success rate by strength of residual coupling. Baseline G success rate decays exponentially as residual coupling increases.}
    \label{fig:sensitivity_residual}
\end{figure}

Fig.~\ref{fig:success_rates} shows worst-case overall success rate, estimated using our heuristic equation~\ref{eq:psuccess}. Note that statistics, such as those from \texttt{qaoa(16)} and \texttt{ising(16)} circuits, are excluded from the analysis due to their estimated success rates being lower than $10^{-4}$. Baseline N is crosstalk-unaware; as a result, crosstalk has detrimental impact on program success rates for any circuit with parallel two-qubit gates on adjacent qubits, as shown in Fig.~\ref{fig:success_rates}. ColorDynamic achieves \emph{comparable performance to Baseline G but with simpler hardware (no tunable couplers)}. Results for Baseline G in Fig.~\ref{fig:success_rates} is a conservative estimate, assuming couplers can be deactivated \emph{perfectly}. We study the effect of residual coupling in Fig.~\ref{fig:sensitivity_residual}.
Compared to Baseline U (with serialization), ColorDynamic consistently outperforms, \emph{achieving 13.3x better success rate on average}. Compared to Baseline S, across all benchmarks, ColorDynamic outperforms static strategies because it is able to exploit program structures and assign frequencies tailored for every layer of the program.

\begin{figure}[t]
    \centering
    \includegraphics[width=0.9\linewidth, trim=0 0cm 0 0]{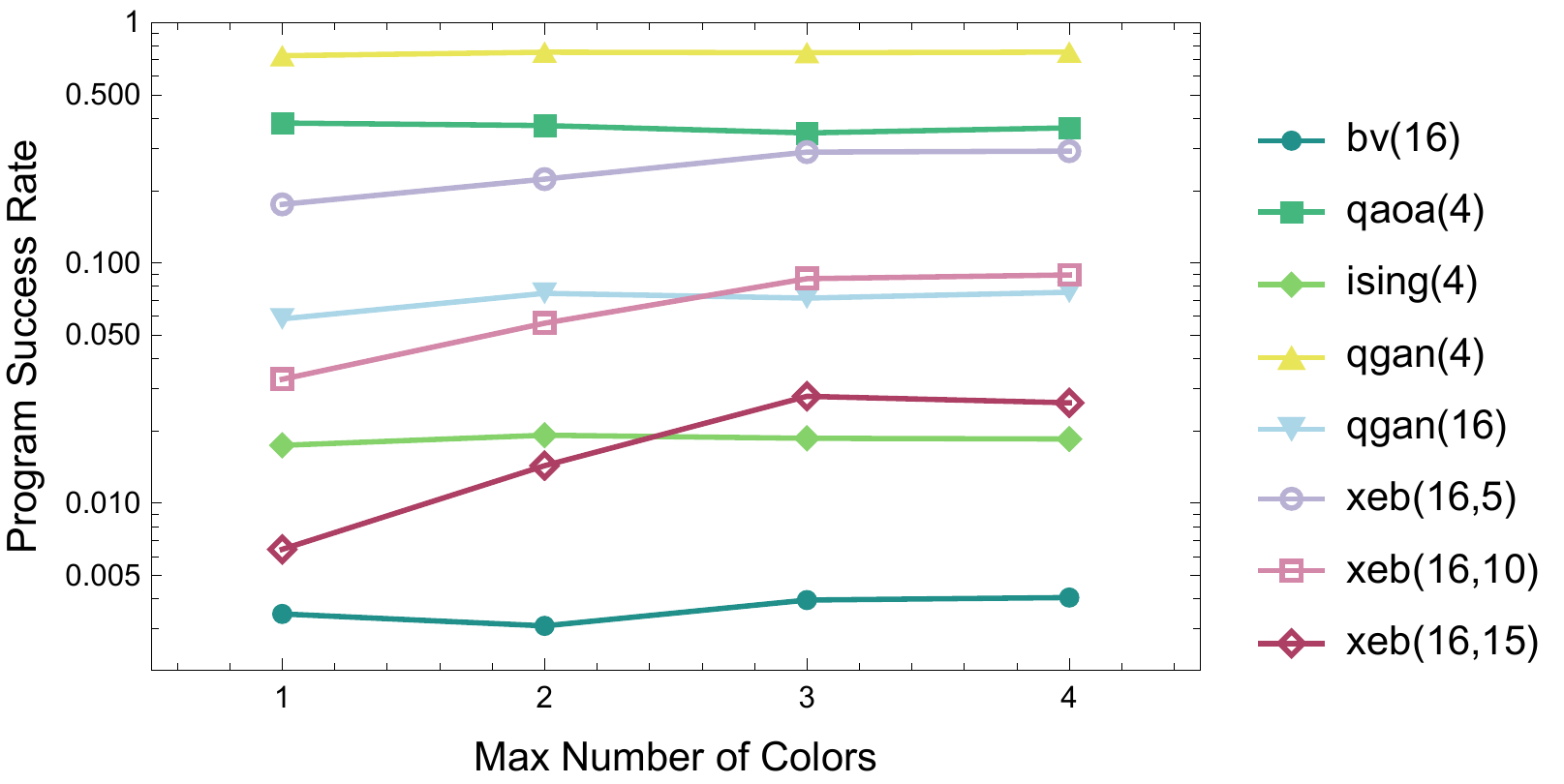}
    \caption{Finding sweet spot of tunability. More than three colors (i.e. frequencies) are typically \emph{unnecessary} for NISQ benchmarks.}
    \label{fig:tunability}
\end{figure}

\subsection{Impact on Serialization}\label{subsec:serial}
Fig.~\ref{fig:depths} compares the resulting program depth and decoherence error across algorithms. Although serialization can effectively prevent gates from crosstalk (commonly adopted such as for IBM's fixed-frequency qubits), it results in deeper circuits (i.e. longer execution time), which consequently implies higher qubit decoherence. Overall, baseline U requires the most amount of serialization. ColorDynamic produces \emph{1.02x average decoherence error}, compared to baseline G, and \emph{0.90x average decoherence error}, compared to baseline U. Lower decoherence error is desirable when executing on NISQ hardware. 

\subsection{Scalability and Complexity}\label{subsec:scalability}

Globally optimizing for the best frequency configuration based on device and program characteristics is challenging; our approach breaks the optimization problem into multiple scalable sub-problems. ColorDynamic keeps the complexity of each sub-problem small, trading off program parallelism for optimization complexity when necessary. In particular, the leading costs stem from coloring of crosstalk graphs and application of SMT solvers. 

\begin{figure*}
    \centering
    \includegraphics[width=0.86\textwidth, trim=0cm 0cm 0 0]{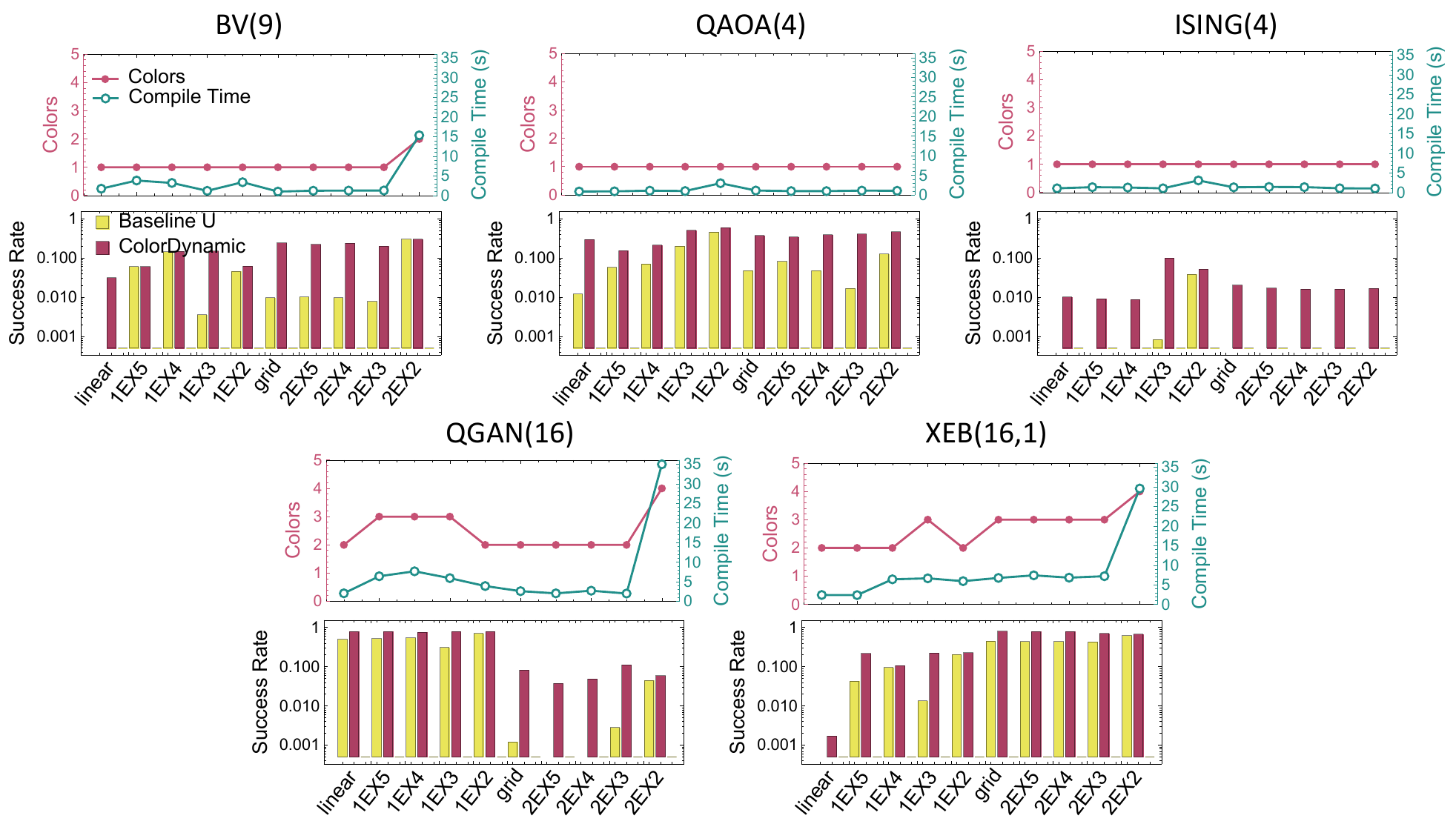}
    \caption{Results on general device connectivity across benchmarks. \textbf{Top:} Number of colors (for interaction frequency) and compilation time of ColorDynamic. \textbf{Bottom:} Log-scale program success rate for Baseline U and ColorDynamic. Denser connectivity from left to right along x-axis. n-EX-k is an $n$-ary express cube \cite{dally1991express} with inserted connections every $k$ nodes.}
    \label{fig:topology}
\end{figure*}

The greedy coloring algorithm takes time polynomially in the graph size, which is kept small thanks to circuit slicing and strategic serialization. The number of variables/constraints in the SMT solver is proportional to the number of colors obtained from coloring; in the next section, we demonstrate that the number of colors remains small. Empirically, we report the \emph{number of colors and compilation time} of ColorDynamic across benchmarks in Fig.~\ref{fig:topology}. Compilation time \emph{remains less than $30$ seconds on systems up to 81 qubits} for a highly parallel benchmark such as \texttt{XEB}.


\subsection{Sensitivity on Tunability}
In ColorDynamic, we can limit the maximum number of colors used for assigning qubit frequencies. To guarantee low crosstalk, fewer colors implies more serialization. In Fig.~\ref{fig:tunability}, we examine the balance between spectral and temporal optimizations, and find the best tunability for each benchmark. In general, we observe optimal operating point at 1 or 2 colors, depending on the initial parallelism of the benchmark. This result has significant hardware implications -- such program-specific optimization shows that frequency-tunable qubits with 2 frequency sweet spots are good candidates for near-term algorithms, hence building qubits with more sweet spots will only give diminishing returns.

\subsection{Gmon's Sensitivity to Residual Coupling}

In our evaluation, Baseline G conservatively assumes that coupling can be (de)activated \emph{perfectly}. In practice, tuning couplers increases sensitivity to control noises. In Fig.~\ref{fig:sensitivity_residual}, we demonstrate how the performance degrade exponentially as residual coupling increases. Such exponential decay in performance motivates the necessity of strategic frequency tuning for tunable qubit and coupler architectures.

\subsection{General Device Connectivity}\label{subsec:connectivity}
To demonstrate the general applicability of our algorithm with respect to device connectivity, we perform a systematic study shown in Fig.~\ref{fig:topology}. Denser connectivity for superconducting device is challenging \cite{kjaergaard2020superconducting}, due to limitations such as coupling and addressing qubits. As such, we target a class of connectivity graphs with increasing density while incurring minimal wiring overhead, namely the ``express cubes'' \cite{dally1991express} designed for interconnection networks. In particular, we augment an increasing number of connections to a 1-D linear path and a 2-D grid, denoted as 1EX-k and 2EX-k graphs respectively, where $k$ stands for inserting a connection every $k$ nodes\cite{dally1991express}.

ColorDynamic consistently \emph{improves program success rate by 3.97x in geometric mean} across all benchmarks, compared to baseline U. Depending on applications, best performance is usually found on connectivity not too sparse or denser than grid. Compilation time of ColorDynamic is kept low ($\sim10$ seconds) in practice, because the number of colors remains small, as argued in Section~\ref{subsec:scalability} and Fig.~\ref{fig:tunability}. Empirically, we see some increase in the extreme cases with unrealistically dense connectivity, but still within a desirable range.

%% file: validation.tex




%% file: conclusion.tex
\section{Conclusion}\label{sec:conclusion}
In this work, we introduce a systematic approach to software mitigation of crosstalk due to frequency crowding.
Our approach allows fixed coupler architectures to compete with tunable coupler architectures in reliability, potentially simplifying the fabrication of quantum machines. The general applicability of our algorithm with respect to device connectivity also motivates potential paths forward in terms of hardware connectivity design. One extension to our work is to apply the methodology of ColorDynamic to guide both qubit tuning and coupler tuning. In fact, the methodology is extensible to any quantum architectures with tunable qubits; it solves a generic calibration problem for isolating or interacting qubits. Finally, complementing Gmon architecture with ColorDynamic optimization would also be a natural extension. 

The compilation and simulation software used in this paper is open-sourced and available on GitHub \cite{git2020fastsc}.


%% file: Append.tex
\section{Example Idle and Interaction Frequencies by C\textsc{olor}D\textsc{ynamic}} \label{sec:example_color}

\begin{figure}[t]
    \centering
    \includegraphics[width=0.42\textwidth, trim=0 0cm 0 0]{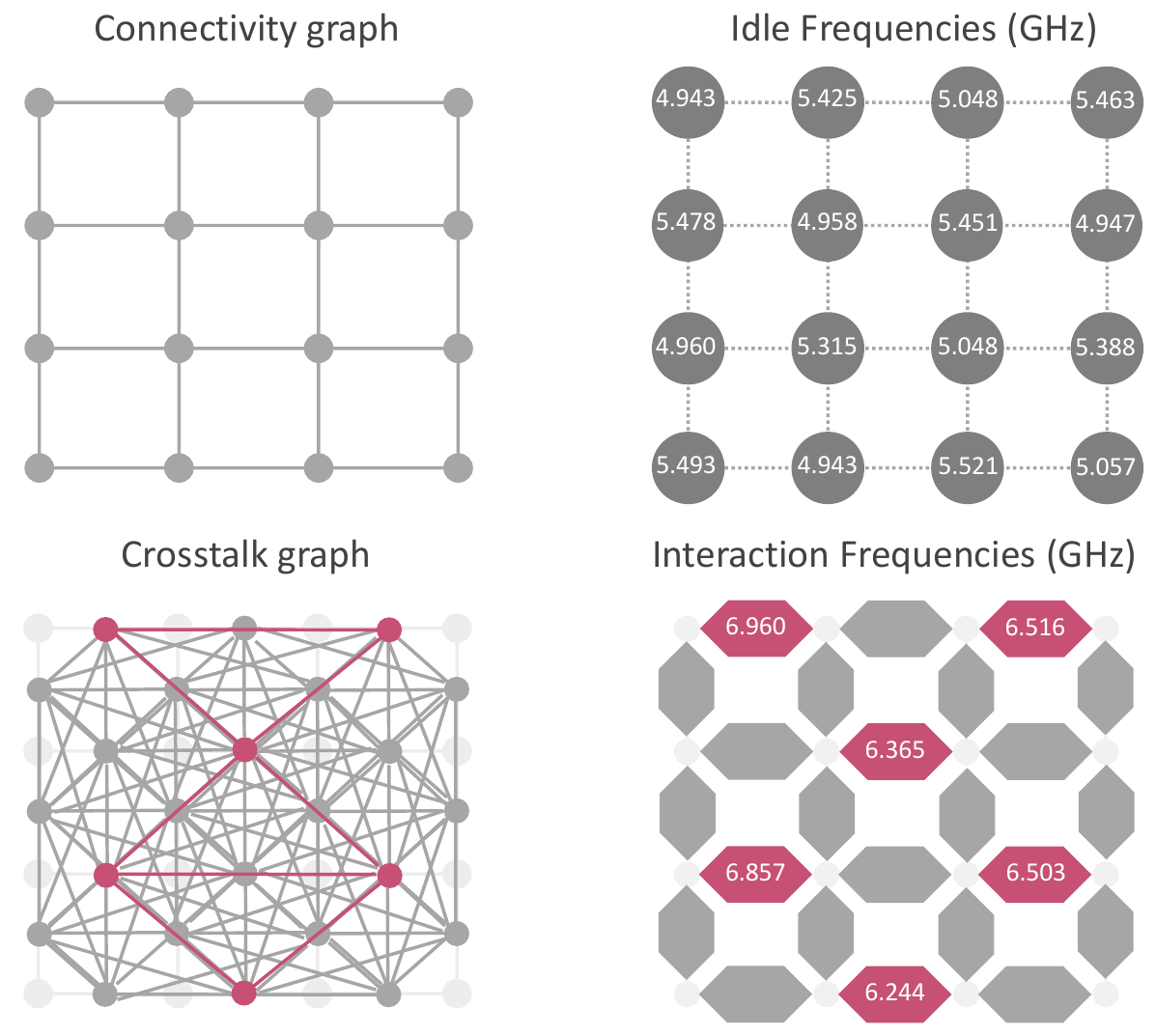}
    \caption{Example qubit frequencies, $\omega_{01}$. \textbf{Top:} the connectivity graph $G_c$ of a $4\times 4$ qubit mesh on the left, and the resulting idle frequencies by coloring $G_c$ on the right. \textbf{Bottom:} the crosstalk graph $G_x$ on the left, and the resulting interaction frequencies for the subgraph of $G_x$ highlighted in red on the right.}
    \label{fig:augmented_line_graph}
\end{figure}
\begin{figure}[t]
    \centering
    \includegraphics[width=0.9\linewidth, trim=0 0cm 0 0]{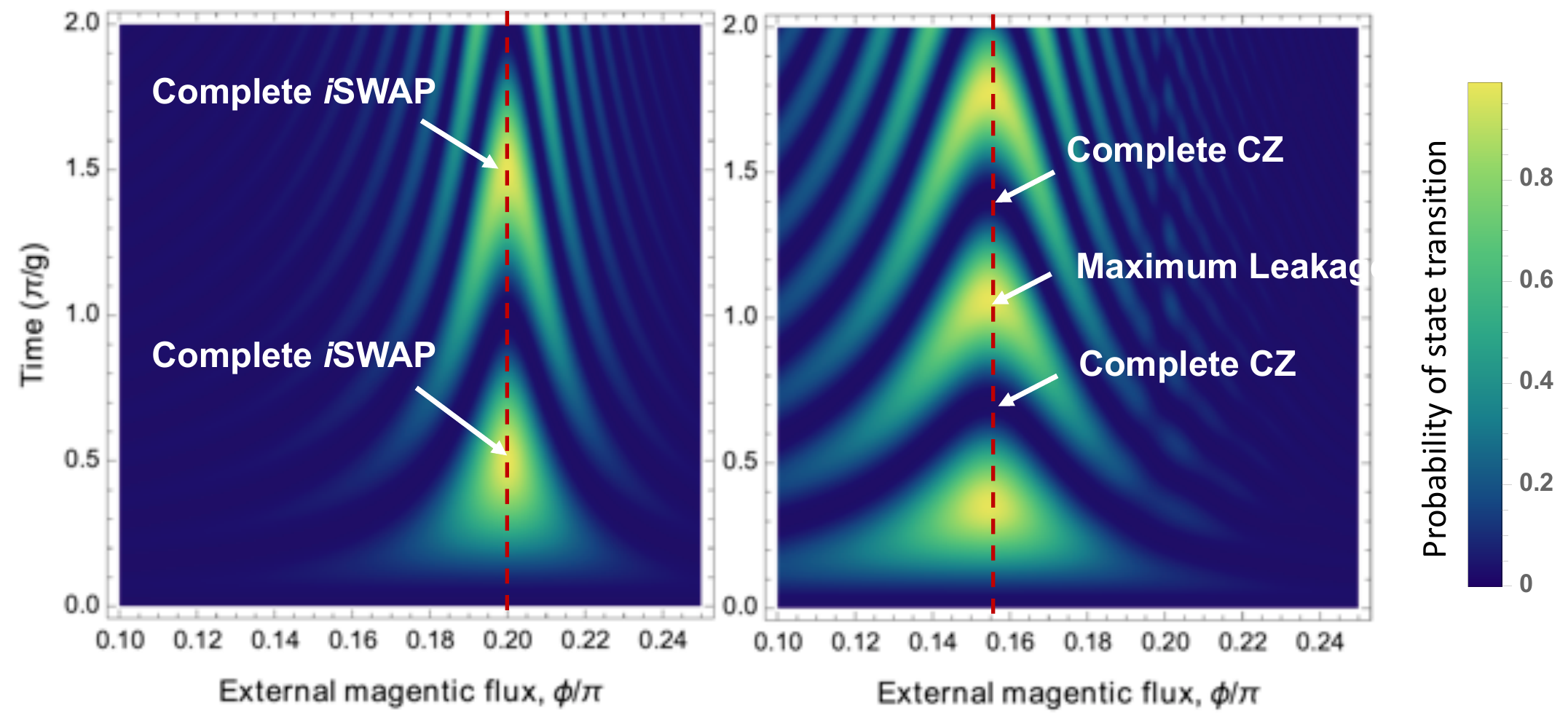}
    \caption{\textbf{Left:} Probability of state transition between $\ket{01}$ and $\ket{10}$, as a function of external magnetic flux and time. \textbf{Right:} Probability of state transition between $\ket{11}$ and $\ket{20}$, as a function of external magnetic flux and time.}
    \label{fig:iswapcz}
\end{figure}

This section provides a concrete example of the idle and interaction frequencies for a $4\times 4$ qubit systems, resulting from the proposed ColorDynamic algorithm, as shown in Fig.~\ref{fig:augmented_line_graph}. Notably, the idle frequencies are assigned in a checker board pattern of high and low values to avoid crosstalk with nearest neighbors. The interaction frequencies are assigned to the qubits performing simultaneous two-qubit gates in one of the time-steps of the \texttt{xeb(16, p)} circuit \cite{arute2019quantum}. Frequencies are optimized by subgraph coloring and SMT solvers. Each asymmetric transmon qubit has two sweet spots, as shown in Fig.~\ref{fig:transmon}. As such, we keep the idle frequencies close to the low sweet spot near $5$ GHz, and the interaction frequencies close to the high sweet spot near $7$ GHz.


\section{Gate Errors due to Crosstalk}\label{sec:append_gates}
We continue to elaborate on our heuristic noise model for estimating gate error $\epsilon_g(\omega, t)$, following Section~\ref{sec:bg} and \ref{sec:eval}. For frequency-tunable transmon qubits, two-qubit gates are accomplished via resonance. Depending on the energy levels that the resonance occurs, we can implement $i$\texttt{SWAP} and \texttt{CZ} gates. In Fig.~\ref{fig:iswapcz}, we plot the probability of state transitions as we tune the local magnetic flux of one of the qubit (along x-axis) and as the time spent on that operating point is increased (along y-axis). Let $\delta\omega = |\omega_A - \omega_B|$ be the frequency difference of two adjacent qubits, with residual coupling strength \cite{krantz2019quantum}: 
\begin{align}
g'(\delta\omega) = \frac{g_0^2}{\hbar^2\delta\omega},
\end{align}
as shown in Fig.~\ref{fig:coupling}, where $g_0$ depends on the effective coupling capacitance $C_{qq}$. The coupling strength determines how fast and strong the state transitions undergo. 
When brought on resonance, the two states $\ket{01}$ and $\ket{10}$ will undergo Rabi oscillation, giving rise to a periodic exchange of energy population. The transition probability is $\Pr[t] = \sin{(gt)}^2$, where $g$ is the coupling strength. Following \cite{barends2019diabatic}, the crosstalk error (for idle qubits) is
\begin{align}
    \epsilon_g(\delta\omega,t) = 1-\sin{(g'(\delta\omega)t)}^2. \label{eq:eg}
\end{align}

For $i$\texttt{SWAP} gate operations, we want a complete exchange of population, predicted at $t=\frac{\pi}{2g}$. We note that for $t=\frac{\pi}{4g}$, it results in another important operation relevant to this work, the $\sqrt{i\texttt{SWAP}}$ gate. The \texttt{CZ} operation is implemented by resonance of $\ket{11}$ and $\ket{20}$. Due to the higher photon number, the coupling strength is scaled by a constant factor, $\sqrt{2}g$. A complete \texttt{CZ} happens when exchanged from $\ket{11}$ to $\ket{20}$, and back to $\ket{11}$, in other words, \texttt{CZ} gate time is $t = \frac{\pi}{\sqrt{2}g}$. 







\section{Overhead of Dynamic Tuning}\label{sec:tuning}
Our algorithm relies on dynamically changing qubit frequencies via an external magnetic flux (Fig.~\ref{fig:transmon}). Our simulation analysis has taken both the \emph{time and error overheads} into account, including flux control noise. Flux tuning has been experimentally demonstrated in fast gate implementations and system calibrations \cite{kjaergaard2020superconducting}. The time overhead of flux tuning is only a fraction of quantum gate. How fast the pulses are changed is parametrically controlled; state-of-the-art controls show accurate, fast tuning (within 2 ns) \cite{rol2019fast}, giving rise to fast single-qubit flux (\texttt{Rz}) gate and two-qubit $i$\texttt{SWAP} and \texttt{CZ} gates (around 50 ns) with $>99.5\%$ fidelity \cite{kjaergaard2020quantum}, compared to a two-qubit \texttt{CR} gate (around 160 ns \cite{sheldon2016procedure}) on fixed-frequency qubit architectures. Control noises and pulse distortions are indeed present in current tunable systems \cite{arute2019quantum}; techniques such as frequency sweet-spots (Fig.~\ref{fig:transmon}) and limiting number of colors (Fig.~\ref{fig:tunability}) are designed to mitigate them.